\documentclass[conference,a4paper,10pt]{IEEEtran}

\usepackage{amsmath, graphics,amssymb,epsfig,subfigure,color,amsthm,cite}

\newtheorem{theorem}{Theorem}

\newtheorem{lemma}{Lemma}

\begin{document}

\sloppy

%\title{IF-OFDM:\\Everyone Gets Half the Cake With Completely No CSIT in the $K$-User Interference Channel with ISI}

\title{Interference-Free OFDM:\\Rethinking OFDM for Interference Networks with Inter-Symbol Interference}

\author{
  \IEEEauthorblockN{Namyoon Lee\\ The Department of Electrical Engineering, POSTECH, \\Pohang, Gyeongbuk, Korea 37673  \\nylee@postech.ac.kr}}
%\thanks{N. Lee  is with the Department of Electrical Engineering, POSTECH, Pohang, Gyeongbuk, Korea 37673 (e-mail:nylee@postech.ac.kr). }
%}

%% Create the title:
\maketitle

\begin{abstract}
This paper considers a $K$-user single-input-single-output interference channel with inter-symbol interference (ISI), in which the channel coefficients are assumed to be linear time-invariant with finite-length impulse response. The primary finding of this paper is that, with no channel state information at a transmitter (CSIT), the sum-spectral efficiency can be made to scale linearly with $K$, provided that the desired links have longer impulse response than do the interfering links. This linear gain is achieved by a novel multi-carrier communication scheme which we call \textit{interference-free orthogonal frequency division multiplexing (IF-OFDM)}. Furthermore, when a transmitter is able to learn CSIT from its paired receiver only, a higher sum-spectral efficiency can be achieved by a two-stage transmission method that concatenates IF-OFDM and vector coding based on singular value decomposition with water-filling power allocation. A major implication of the derived results is that separate encoding across subcarriers per link is sufficient to linearly increase the sum-spectral efficiency with $K$ in the interference channel with ISI. Simulation results support this claim.

\end{abstract}

%\begin{keywords}
%Compressed sensing, Lattice codes,
%\end{keywords}

%%%%%%%%%%%%%%%%%%%%%%%%%%%%%%%%%
%%%%%%%%%%%%%%%%%%%%%%%%%%%%%%%%%
\section{Introduction}
 In multi-user wideband wireless systems, two fundamental obstacles limit the gains in spectral efficiency:
\begin{itemize}

\item Inter-symbol interference (ISI): In a wideband communication system that uses a band-limited signal, ISI is fundamentally unavoidable when the transmission bandwidth $W$ exceeds the coherence bandwidth $W_{\rm c}$ of the channel. ISI distorts the signal between subsequent transmitted symbols; this effect limits the spectral efficiency of a wideband communication system.

\item Inter-user interference (IUI): In a multi-cell communication system that uses universal frequency reuse per cell, IUI (or inter-cell interference) is also inevitable. IUI limits the spectral efficiency of the multi-cell system, because it reduces signal-to-interference plus noise ratio (SINR) of the users.
\end{itemize}

%In the multi-user scenario, inaccurate CSIT
%causes the inevitable inter-user interference (IUI).

%The problem of successfully mitigating these two types of interference is a long-standing but unsolving problem.  
% OFDM:66, Weinstein:71, Hirt_Massey:88,Verdu:93,Tse:97,Wong:99, Rhee_Cioffi:00,Jindal_Goldsmith:04,Yu:07,Cioffi:07

The problem of mitigating both ISI and IUI simultaneously is challenging in general. The common approach has been to deal with the problem of mitigating them separately. For example, orthogonal frequency division multiplexing (OFDM)\cite{OFDM:66, Weinstein:71,Bingham:90} is a well-known method that successfully removes ISI in single-user wideband communication systems. The key principle of the OFDM is to change the linear convolution relation between  input signals and the impulse response of the ISI channel to be the circular convolution between them by adding an appropriate cyclic prefix. Then, using the inverse discrete Fourier transform (IDFT) as transmit eigen-beamforming, and the discrete Fourier transform (DFT) as receive eigen-beamforming, the ISI channel is decomposed into a set of orthogonal subchannels (subcarriers). Essentially, this channel decomposition creates multiple parallel Gaussian channels, each experiencing ISI-free narrowband flat fading. It has shown that this strategy achieves the capacity of the ISI channel with a simple water-filling power allocation \cite{Hirt_Massey:88} in an asymptotic sense, as the number of subcarriers tends to infinity.

The concept of this channel decomposition creating parallel subchannels has been extensively extended to multi-user scenarios \cite{Verdu:93,Tse:97,Wong:99, Rhee_Cioffi:00,Jindal_Goldsmith:04,Yu:07,Cioffi:07} by incorporating IUI. For instance, by allowing two transmitters to share an OFDM symbol to communicate with a receiver, the capacity region for such multiple access channel (MAC) with ISI was characterized in \cite{Verdu:93}. Similarly, the capacity for the two-user interference channel with ISI is also established, especially when IUI is strong \cite{Cioffi:07}. The common achievable scheme revealing the capacity results is the use of separate coding over each subchannel with multiuser water-filling algorithms  based on game-theoretic optimization. For the general case of the $K$-user interference channel with ISI, which can be transformed into the $K$-user parallel Gaussian interference channel by OFDM \cite{Weinstein:71}, the capacity remains open. It has shown in \cite{Cadambe_Jafar:08} that the optimal sum degrees of freedom (sum-DoF) of the parallel $K$-user interference channel is $\frac{K}{2}$, assuming the channel coefficients are independent across subchannels. The key to this sum-DoF characterization is a novel IUI management method, referred to as \textit{interference alignment} (IA). In contrast to the Gaussian point-to-point, multiple access, and broadcast channels with ISI in which separate coding is optimal \cite{Hirt_Massey:88,Verdu:93,Tse:97}, joint encoding across subchannels has shown to provide a significant spectral efficiency gain over separate coding when signal-to-noise ratio (SNR) is high for the parallel Gaussian interference channel \cite{Cadambe_Jafar:09}. 

Despite the significant gains, IA requires global and instantaneous channel state information at transmitter (CSIT), so IA is too difficult to be used as a practical interference management solution in many multi-user wideband wireless systems. 

In the absence of knowledge about CSIT, the sum-DoF of numerous wireless networks is the same as what is achievable by time-division multiple access (TDMA) among users \cite{Jafar:05, Guo:12, Huang:12, Vaze:12}. Recent research on IA has made progress towards using limited CSIT to realize the gains of IA in practical wireless systems \cite{JafarBIA,Wang,Maddah-Ali:12,Namyoon_STIA1,Namyoon_STIA,Tandon:13,Chen_Elia:13
,Jafar_Index_coding}. Blind interference alignment \cite{JafarBIA} is a representative technique that uses the knowledge of channel coherence patterns at the transmitters instead of instantaneous channel coefficients themselves. However, diversity of coherence intervals among different receivers can naturally occur in wireless channels (or can be induced artificially using reconfigurable antennas  \cite{Wang}); by exploiting this diversity, one can achieve the same DoF in an SISO interference channel as that attained with perfect knowledge of CSIT \cite{JafarBIA,Wang}. The impediment to implementing these techniques is that to construct a particular super-symbol structure, the cooperative transmitters must know the channel coherence pattern information (second-order channel statistics) from all receivers. Due to insertion loss in RF circuits, the use of switching-reconfigurable antennas \cite{Wang} is also not preferable to current wireless systems.

\begin{figure}
\centering \vspace{-0.1cm}
\includegraphics[width=3.5in]{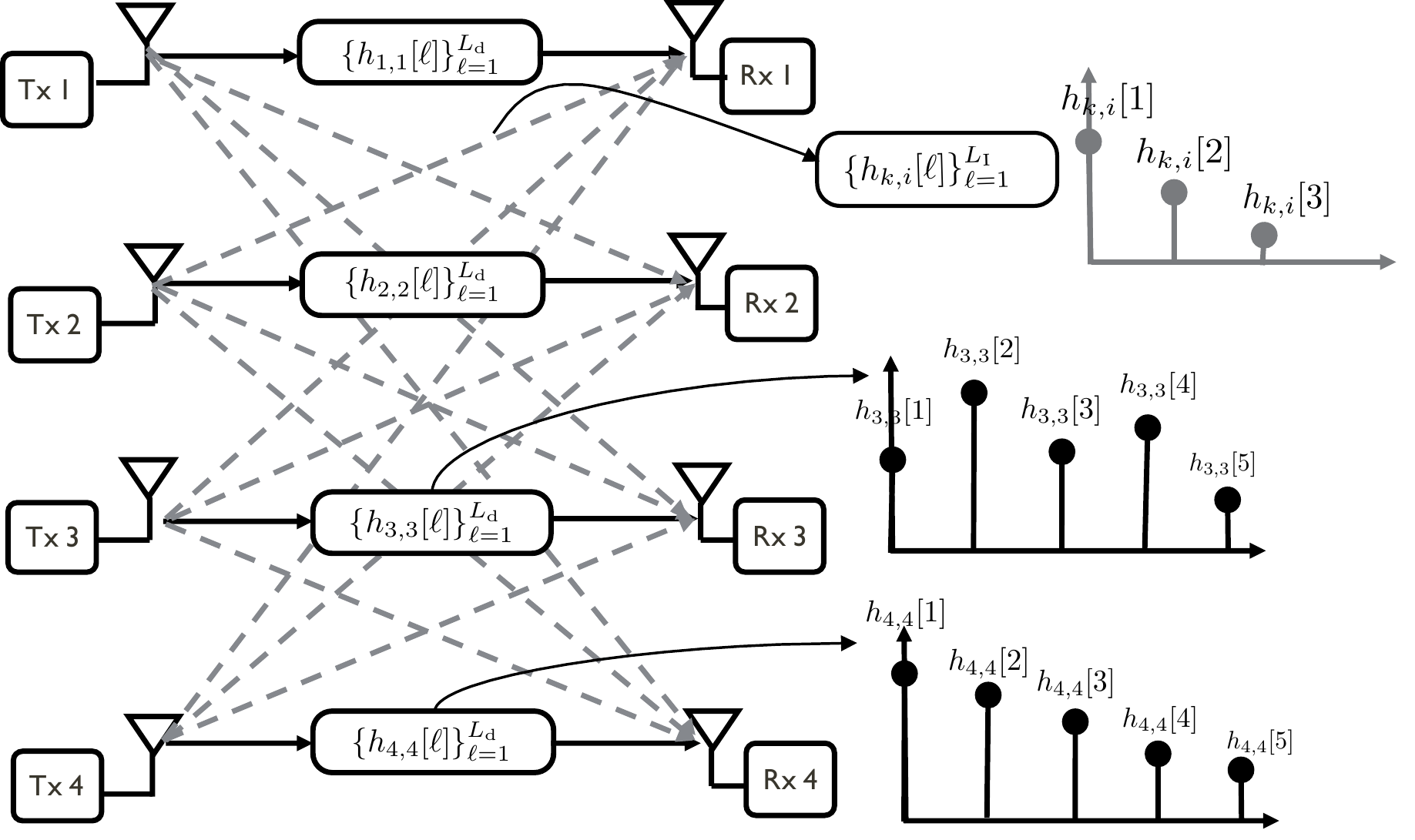} \vspace{-0.1cm}\caption{The system model for the $K$-user interference channel with ISI when $K=4$, $L_{k,k}=L_{\rm D}=5$, and $L_{k,i}=L_{\rm I}=3$.} \label{fig:1} \vspace{-0.3cm}
\end{figure}

In this paper, we consider a single-input-single-output (SISO) $K$-user interference channel with ISI as illustrated in Fig. \ref{fig:1}. It is important to notice that this channel model is not necessarily  equivalent to the parallel (or multi-carrier) Gaussian interference channel. One can transform the interference channel with ISI to the corresponding multi-carrier interference channel by using the decomposition approach by OFDM \cite{Hirt_Massey:88}, but not necessarily vice versa. Throughout the paper, we focus on assumptions that the channel coefficients are linear time-invariant (LTI) with finite-length impulse response. Most importantly, we assume that transmitters have completely lack CSIT; i.e., that they do not even know channel coherence patterns that are essential for implementing the conventional blind interference alignment \cite{JafarBIA}. In this setting, \textit{ the fundamental question we address in this paper is whether the sum-spectral efficiency can be made to scale linearly with $K$, even with complete lack of CSIT.}

Our major contribution of this paper is to show that the answer to this question is affirmative in some ISI conditions. More precisely, we demonstrate that the sum-DoF of the $K$-user interference channel with ISI is
\begin{align}
d^{\rm IC}_{\Sigma} = \max\left\{ \sum_{k\in\mathcal{K}}\frac{(L_{k,k}-L_{\rm I})^{+}}{(\max\{L_{\rm I}, 2(L_{\rm D} -L_{\rm I}) \}+L_{\rm I}-1)}, 1 \right\}, \nonumber
\end{align} 
where $L_{\rm D}=\max_{k}\{L_{k,k}\}$ and $L_{\rm I}=\max_{k}\max{i\neq k}\{L_{k,i}\}$ are the maximum impulse response lengths of the desired and interfering links, respectively. For example, when $L_{k,k}=L_{\rm D}=2$ and $L_{k,i}=L_{\rm I}=1$, $\frac{K}{2}$ sum-DoF can be achieved in a channel that has ISI. This is a remarkable result because it has shown that the optimal sum-DoF of the multi-carrier $K$-user SISO interference channel, $\frac{K}{2}$, is achievable only when global and perfect CSIT is available \cite{Cadambe_Jafar:08} or the specialized channel realizations occur  \cite{JafarBIA,Nazer}. Our result, however, shows that, with complete lack of CSIT, the sum-DoF increases linearly with $K$ no matter how the channel coefficients realized, provided that $L_{k,k}>L_{\rm I}$.

To show the achievability of this result, we introduce a novel communication method, called \textit{interference-free OFDM (IF-OFDM)}. The principal idea of IF-OFDM is to convert the effective channel matrices of all interfering wireless links to circulant channel matrices, while keeping the non-circulant structure for the channel matrix of the desired link by judiciously choosing the cyclic prefix size. Using the fact that column vectors of an IDFT matrix are the common right eigenvectors of circulant matrices, each transmitter sends its information along a set of linearly independent column vectors of the IDFT matrix. This strategy essentially allows aligning all the interference within the same subspace at each receiver, while the desired signals remain out of the interference subspace. By projecting the received signals onto the subspace that is orthogonal to the interference subspace, all IUI are removed perfectly. Once IUI is eliminated, each receiver decodes the desired information symbols by removing the remaining inter-subcarrier interference.

%By exploiting this relativity of the channel structures, each transmitter sends multiple data symbols using the same eigen-beamforming based on IDFT, which essentially require completely no CSIT. Since the effective interference channels are circulant, this eigen-beamforming makes all interference signals align in the same subspace regardless of the realization of channel coefficients. At the same time, it also allows to preserve the desired signals out of the interference subspace, as the effective channel matrix was transformed to a non-circulant matrix. 

By using this method, we also characterize an achievable sum-spectral efficiency of the $K$-user interference channel with ISI under two CSIT assumptions on the desired link. For the case of completely absent CSIT, we obtain a closed-form expression of the sum-spectral efficiency that achieved by the concatenation of two transmission methods: 1) IF-OFDM for mitigating IUI and 2) zero-forcing based successive interference cancellation (ZF-SIC) to remove inter-subcarrier interference. Furthermore, when a transmitter can learn CSIT from its paired receiver only, we also yield a characterization of an achievable sum-spectral efficiency of the $K$-user interference channel with ISI. To derive this result, a two-stage precoding method is presented, which essentially concatenates IF-OFDM to mitigate IUI and singular value decomposition (SVD) based vector coding with water-filling power allocation to reject inter-carrier interference. One remarkable feature of the proposed method is the use of separate encoding across subchannels; this implies that a data symbol of each transmitter is sent through a subcarrier. This fact is both practically relevant and theoretically interesting, because it reveals that separate encoding achieves a significant spectral efficiency gain in a multi-carrier interference channel.

%
%The rest of the paper is organized as follows. In Section II,
%a system model of the SISO $K$-user interference channel with ISI is described. We illustrate the key idea of the proposed IF-OFDM through a simple example in Section III. In Section IV, the main theorem giving the sum-DoF of the SISO $K$-user interference channel with ISI is presented. In Section V, we provide the characterization of achievable sum-spectral efficiency under two assumptions on the desired link's CSIT. In Section VI, we provide some discussion to apply the proposed method under practical ISI conditions. The paper concludes with future directions in Section VII. 
%

The rest of the paper is organized as follows. We describe a system model of the SISO $K$-user interference channel with ISI in Section II. We illustrate the key idea of the proposed IF-OFDM through simple examples in Section III. We present the main theorem giving the sum-DoF of the SISO $K$-user interference channel with ISI in Section IV. Then, in Section V, we provide the characterization of achievable sum-spectral efficiency under two assumptions on the desired link’s CSIT. We also provide some discussion to apply applications of the proposed method under practical ISI conditions in Section VI. The paper concludes with future directions in Section VII.
%
%Throughout this paper, transpose and conjugate transpose of a matrix ${\bf X}$ are represented by ${\bf X}^{\top}$ and 
%${\bf X}^{H}$,respectively. In addition, $\mathbb{C}$ and $\mathbb{R}$ indicates a complex and real value.

%%%%%%%%%%%%%%%%%%%%%%%%%%%%
%%%%%%%%%%%%%%%%%%%%%%%%%%%%
%%%%%%%%%%%%%%%%%%%%%%%%%%%%

\section{Signal Model}
We consider a $K$-user SISO interference channel with ISI. As illustrated in Fig. \ref{fig:1}, transmitter $k\in\mathcal{K}\triangleq\{1,2,\ldots,K\}$ uses the shared spectrum to communicate with its associated receiver $k\in\mathcal{K}$. We assume that all transmitters and receivers are equipped with a single antenna.  

Let $x_k[n]$ be the transmitting signal of transmitter $k\in\mathcal{K}$ at time slot $n\in\mathbb{Z}^+$ with an average power constraint $\mathbb{E}[|x_k[n]|^2]=P$. We also denote the channel impulse response from transmitter $k$ to receiver $j$ by $\{h_{k,i}[\ell]\}_{\ell=1}^{L_{k,i}}$, in which $L_{k,i}$ is the effective number of multi-paths between transmitter $k$ and receiver $j$. Then, the received signal of receiver $k \in \mathcal{K}$ at time slot $n$ is
\begin{align}
y_{k}[n]=\sum_{i=1}^K\sum_{\ell=1}^{L_{k,i}}h_{k,i}[\ell]x_{i}[n-\ell+1]+ z_k[n], \label{eq:received_system_model}
\end{align}
where $z_k[n]$ is the noise at receiver $k$ during time slot $n$. Here, the number of channel-taps of the wireless channel from transmitter $i$ to receiver $k$ is typically defined as
\begin{align}
L_{k,i} \triangleq\left\lceil T^{\rm D}_{k,i}W\right\rceil,
\end{align}
where $T^{\rm D}_{k,i}$ is the delay spread of the channel from transmitter $i$ to receiver $k$, $W$ is the transmission bandwidth of a signal, and $\lceil x\rceil $ is the smallest integer greater than or equal to $x$. For example, if symbol duration $T_{\rm s}=\frac{1}{W}$ is larger than delay spread $T^{\rm D}_{k,i}$, i.e.,  $L_{k,i}=1$ for all $i,k\in\mathcal{K}$, this model is equivalent to the $K$-user SISO frequency-flat (narrowband) interference channel.

In our model, we assume that fading coefficients of $\{h_{k,i}[\ell]\}$ are time-invariant for a communication duration, i.e., a block-fading assumption. In each block, all the channel coefficients $\{h_{k,i}[\ell]\}$ are drawn from a continuous distribution. In a rich-scattering propagation environment, for instance, each channel coefficient can be modeled as an independent complex Gaussian random variable with zero-mean and variance that decays exponentially as the number of taps increases, i.e., $h_{k,i}[\ell]\simeq \mathcal{CN}(0,e^{-\beta(\ell-1)})$, where $\beta$ is the constant that determines the rate of channel power reduction.
Furthermore, $z_k[n]$ are assumed to be independent and identically-distributed (IID) complex Gaussian random variables with zero mean and variance $\sigma^2$, i.e., $\mathcal{CN}(0,\sigma^2)$. Throughout the paper, we assume that all transmitters have no CSIT, and that each receiver has only knowledge of the channel impulse response between itself and the paired transmitter; i.e., no knowledge of channel state information at receiver (CSIR) associated with other receivers. This is the minimum channel knowledge to perform coherent detection at receivers.

% Since the channel is assumed to be frequency-selective, wireless channel inherently provides frequency diversity. Whereas, there is no time-diversity in our channel model assumption. 

Transmitter $k$ sends an independent message  $m_{k}$ to  intended receiver $k$ at rate $R_{k}(P)=\frac{\log_2|m_{k}|}{M}$ for $k \in\mathcal{K}$. Then, it is said that rate $R_{k}(P)$ is achievable if receiver $k$ can decode the desired message with an error probability that is arbitrarily small for sufficient channel uses $M$. The sum-DoF that characterizes an approximate sum-spectral efficiency at high \textrm{SNR} is defined as \begin{align}
d_{\Sigma}\!&\!=\lim_{{P}\rightarrow \infty}\frac{\sum_{k=1}^{K}R_{k}\left( P\right)}{\log\left({P}\right)}.
\end{align}
We aim at using the sum-DoF metric in this paper 
to provide a clear understanding of our proposed transmission method in Section III and IV. In Section V, we will characterize the expressions of the sum-spectral efficiency when the proposed method is applied.

 %
%suitably captures the signal interactions by deemphasizing the effects of noise, thereby  of the sum-capacity for sophisticated networks.The achievable rate
%region $\mathcal{R}(P)$ is defined as the set of
%achievable spectral efficiency tuples $\mathcal{R}(P)=\left( R_1(P),R_2(P),\ldots, R_K(P)\right)$. 
% 

\section{IF-OFDM}
In this section, we present the key idea behind IF-OFDM starting with some example cases. Gaining insights
from this section, we provide our main result for  general scenarios in Section IV.

 {\bf Example 1:} Suppose $L_{k,i}=L_{\rm I}=2$ and $L_{k,k}=L_{\rm D}=3$ for $i,k\in\mathcal{K}$. In this example case, our goal is to show that each transmitter is able to send one data symbol to the associated receiver using four time slots, i.e., $d_{\Sigma}=\frac{K}{4}$. The key to demonstrating this result is the proposed transmission method, referred as to \textit{IF-OFDM}. The principal idea of IF-OFDM is to create a circulant channel structure for all interfering wireless links, while keeping the non-circulant structure for the desired wireless link. Because of this relativity of the channel structures between the interfering and the desired links, very simple precoding can be used to align all interference signals in a subspace while preserving the desired signals in a different subspace.

 %
%
%
%the difference between the number of channel-taps of the desired link and that of interfering link in order to create \textit{relativity of alignment}. 
%
% a subset of column vectors of a DFT (or IDFT) matrix as 
%
% (even without using knowledge of particular channel coherence patterns) 

Let ${\bf f}_1=\frac{1}{\sqrt{2}}\left[1, 1\right]^{\top}$ be the beamforming vector carrying information symbol $s_k$ for $k\in\mathcal{K}$. This beamforming vector will be used for all transmitters. The transmitted signals of transmitter $k$ during two time slots can be interpreted as a simple repetition transmission method, namely,
\begin{align}
{\bf x}_k=\left[%
\begin{array}{c}
{x}_{k}[1]    \\
{x}_{k}[2]  \\
\end{array}%
\right] = {\bf f}_1s_k,
\end{align}
where $k\in\mathcal{K}$.
Then, from \eqref{eq:received_system_model} the received signals of receiver $k$ during four time slots are given by
\begin{align}
 \underbrace{\left[%
\begin{array}{c}
{y}_{k}[1]   \\
{y}_{k}[2]  \\
{y}_{k}[3]  \\
{y}_{k}[4]  \\
\end{array}%
\right]}_{{\bf y}_k} \!&\!= \underbrace{\left[%
\begin{array}{cc}
{h}_{k,k}[1]  \!&\! 0 \\
{h}_{k,k}[2]  \!&\! {h}_{k,k}[1] \\
{h}_{k,k}[3]  \!&\! {h}_{k,k}[2]  \\
0  \!&\! {h}_{k,k}[3]  \\
\end{array}%
\right]}_{{\bf H}_{k,k}}  \left[%
\begin{array}{c}
{x}_{k}[1]   \\
{x}_{k}[2]  \\
\end{array}%
\right] \nonumber \\\!&\!+ \sum_{i\neq k}^K\underbrace{\left[\!\!%
\begin{array}{cc}
{h}_{k,i}[1]  \!&\! 0 \\
{h}_{k,i}[2]  \!&\! {h}_{k,i}[1] \\
0  \!&\! {h}_{k,i}[2]  \\
0  \!&\! 0  \\
\end{array}%
\!\!\right] }_{{\bf H}_{k,i}}\!\!\! \left[\!\!%
\begin{array}{c}
{x}_{i}[1]   \\
{x}_{i}[2]  \\
\end{array}%
\!\!\right] \!\!+\!\! \left[\!\!%
\begin{array}{c}
{z}_{k}[1]   \\
{z}_{k}[2]  \\
{z}_{k}[3]  \\
{z}_{k}[4]  \\
\end{array}%
\!\!\right]. \label{eq:example_rx2}
\end{align} 
As mentioned previously, the key idea of IF-OFDM is to 1) make all the channel matrices of interfering links ${\bf H}_{k,i}$ be circulant matrices, 2) while ensuring that desired channel matrix ${\bf H}_{k,k}$ does not hold the circulant matrix structure. Here, we present an idea of how to accomplish these above two conditions.

Let ${\bf \tilde D}_k\in \{0,1\}^{2\times 4}$ be a  combiner matrix that is applied for receiver $k$. In this example, we particularly construct ${\bf \tilde D}_k$ for $k\in\mathcal{K}$ as follows:
\begin{align}
{\bf \tilde D}_k =  \left[%
\begin{array}{cccc}
1 \!&\! 0 \!&\! 1 \!&\! 0  \\
0 \!&\! 1 \!&\! 0 \!&\! 0  \\
\end{array}%
\right].
\end{align}
Applying this linear combiner to ${\bf y}_k$ in \eqref{eq:example_rx2}, we obtain output vector ${\bf \bar y}_{k} ={\bf \tilde D}_k{\bf y}_k$ as
\begin{align}
&\underbrace{\left[\!\!\!%
\begin{array}{c}
{y}_{k}[1]  \!+\! {y}_{k}[3]  \\
{y}_{k}[2]  \\
\end{array}%
\!\!\!\right] }_{ {\bf \bar y}_{k}}
= \!\!\underbrace{\left[\!\!\!%
\begin{array}{cc}
{h}_{k,k}[1]\!+\!{h}_{k,k}[3]    \!\!&\!\! {h}_{k,k}[2]  \\
{h}_{k,k}[2]  \!\!&\!\! {h}_{k,k}[1] \\
\end{array}%
\!\!\!\right]}_{{\bf \bar H}_{k,k}} \!\underbrace{  \left[\!\!\!%
\begin{array}{c}
{x}_{k}[1]   \\
{x}_{k}[2]  \\
\end{array}%
\!\!\!\right]}_{{\bf x}_k} \nonumber \\\!&\!\!\!\!\!\!\!\!\!\!+ \sum_{\ell\neq k}^K\underbrace{\left[\!\!%
\begin{array}{cc}
{h}_{k,i}[1]  \!\!&\!\! {h}_{k,i}[2] \\
{h}_{k,i}[2]  \!\!&\!\! {h}_{k,i}[1] \\
\end{array}%
\!\!\right] }_{{\bf 
\bar H}_{k,i}} \!\underbrace{\left[\!\!\!%
\begin{array}{c}
{x}_{i}[1]   \\
{x}_{i}[2]  \\
\end{array}%
\!\!\!\right]}_{{\bf x}_{i}} \!+\! \underbrace{\left[\!\!\!\!%
\begin{array}{c}
{z}_{k}[1] \!+\!{z}_{k}[3]  \\
{z}_{k}[2]  \\
\end{array}%
\!\!\!\!\right]}_{{\bf \bar z}_k}\!\!. \label{eq:output_combiner}
\end{align}
In \eqref{eq:output_combiner}, all the effective interference channel matrices ${\bf 
\bar H}_{k,i}\in \mathbb{C}^{2\times 2}$ are circulant,  whereas the effective desired matrix ${\bf 
\bar H}_{k,k}\in \mathbb{C}^{2\times 2}$ is not a circulant matrix. It is well-known that a circulant matrix ${\bf  \bar H}_{k,i}$ can be diagonalized by a discrete-time Fourier transform (DFT) matrix that obeys 
\begin{align}
{\bf  \bar H}_{k,i}=\left[%
\begin{array}{cc}
{\bf f}_1   \!&\! {\bf f}_2
\end{array}%
\right] \left[%
\begin{array}{cc}
 \lambda^{1}_{k,i}   \!&\! 0 \\
0 \!&\! \lambda^{2}_{k,i}
\end{array}%
\right]\left[%
\begin{array}{cc}
{\bf f}_1   \!&\! {\bf f}_2
\end{array}%
\right]^H,
\end{align}
where ${\bf f}_1=\frac{1}{\sqrt{2}}[1,1]^{\top}$ and ${\bf f}_2=\frac{1}{\sqrt{2}}[1,-1]^{\top}$ are the eigenvectors, and $\lambda^{p}_{k,i}$ denotes the eigenvalue associated with eigenvector ${\bf f}_p$ for $p=\{1,2\}$. Recall that all transmitters have applied the same linear beamforming vector for sending each data symbol, i.e., ${\bf x}_k={\bf f}_1{ s}_k$ for $k\in\mathcal{K}$. As a result, all interference signals are aligned in the signal direction ${\bf f}_1$, i.e.,
\begin{align}
{\rm span}({\bf 
\bar H}_{k,i}{\bf f}_1) = {\rm span}({\bf 
\bar H}_{k,j}{\bf f}_1),
\end{align}  
for all $j,i \in\mathcal{K}/\{k\}$ and $j\neq i$.
This fact allows us to rewrite ${\bf \bar y}_k$ in \eqref{eq:output_combiner} as follows:
\begin{align}
{\bf \bar y}_k \!&\!= {\bf 
\bar H}_{k,k}{\bf f}_1s_k + \sum_{i\neq k}{\bf 
\bar H}_{k,i}{\bf f}_1s_{i} + {\bf 
\bar z}_{k} \nonumber \\
\!&\!= {\bf 
\bar H}_{k,k}{\bf f}_1s_k + \sum_{i\neq k}\lambda^1_{k,i}{\bf f}_1s_{i} + {\bf 
\bar z}_{k}.
\end{align}
Because ${\bf f}_1$ and ${\bf f}_2$ are mutually orthogonal to each other, we multiply receive beamforming vector ${\bf f}_2^H$ to ${\bf \bar y}_k$; by this process, it possible to eliminate all the aligned interference at receiver $k$ can be eliminated. Accordingly, the interference-free received signal output is given by
\begin{align}
{\tilde y}_k = {\bf f}_2^H{\bf \bar y}_k = {\bf f}_2^H {\bf 
\bar H}_{k,k}{\bf f}_1s_k + {\bf f}_2^H{\bf 
\bar z}_{k}.
\end{align}
Therefore, the spectral efficiency of the $k$th communication link is 
\begin{align}
R_k &= \frac{1}{4} \log_2\left(1+ \frac{2P|{\bf f}_2^H {\bf 
\bar H}_{k,k}{\bf f}_1|^2}{3\sigma^2}\right) \nonumber \\
 &=\log_2\left(1+ \frac{P|h_{k,k}[3]|^2}{6\sigma^2}\right),
\end{align} 
where $\mathbb{E}[|{\bf f}_2^H{\bf 
\bar z}_{k}|^2]=\frac{3}{2}\sigma^2$, $\mathbb{E}[|s_k|^2]=P$, and ${\bf f}_2^H {\bf 
\bar H}_{k,k}{\bf f}_1=\frac{-h_{k,k}[3]}{2}$. Because we have used four time slots to deliver $K$ independent data symbols, the sum-DoF of $\frac{K}{4}$ is achievable, i.e., $d_{\Sigma}=\frac{K}{4}$.

{\bf Example 2:} Suppose $L_{k,k}=L_{\rm D}=3$ and $L_{k,i}=L_{\rm I}=4$. We assume that $h_{k,k}[\ell] \neq 0$ for $\ell\in\{1,2,3\}$ while, $h_{k,i}[1]=h_{k,i}[2]=0$ and $h_{k,i}[\ell ] \neq 0$ for $\ell\in\{3,4\}$. Therefore, $ |{\rm supp} \left( {h}_{k,k}[1], h_{k,k}[2], h_{k,k}[3]\right) |=3 $ and $|{\rm supp} \left( {h}_{k,i}[1], h_{k,i}[2],\ldots, h_{k,i}[4]\right) |=2$, where ${\rm supp}({\bf x})$ denotes a support set of ${\bf x}$. This example corresponds to the scenario in which the arrival-times of the desired signal and interference signals differ due to the propagation delay between the desired and the interfering links. In this example, we will show that each transmitter sends one information symbol by spanning five time slots. Let ${\bf f}_1=\frac{1}{\sqrt{2}}\left[1, ~ 1\right]^{\top}$ be the beamforming vector carrying information symbol $s_k$ for $k\in\mathcal{K}$. Then, from \eqref{eq:received_system_model}, the received signals of receiver $k$ during five time slots are given by
\begin{align}
 \underbrace{\left[%
\begin{array}{c}
{y}_{k}[1]   \\
{y}_{k}[2]  \\
{y}_{k}[3]  \\
{y}_{k}[4]  \\
{y}_{k}[5]  \\
\end{array}%
\right]}_{{\bf y}_k} \!&\!= \underbrace{\left[%
\begin{array}{cc}
{h}_{k,k}[1]  \!&\! 0 \\
{h}_{k,k}[2]  \!&\! {h}_{k,k}[1] \\
{h}_{k,k}[3]  \!&\! {h}_{k,k}[2]  \\
0  \!&\! {h}_{k,k}[3]  \\
0  \!&\! 0  \\
\end{array}%
\right]}_{{\bf H}_{k,k}}  \left[%
\begin{array}{c}
{x}_{k}[1]   \\
{x}_{k}[2]  \\
\end{array}%
\right] \nonumber \\\!&\!+ \sum_{i\neq k}^K\underbrace{\left[\!\!%
\begin{array}{cc}
0  \!&\! 0  \\
0  \!&\! 0  \\
{h}_{k,i}[3]  \!&\! 0 \\
{h}_{k,i}[4]  \!&\! {h}_{k,i}[3] \\
0  \!&\! {h}_{k,i}[4]  \\
\end{array}%
\!\!\right] }_{{\bf H}_{k,i}}\!\!\! \left[\!\!%
\begin{array}{c}
{x}_{i}[1]   \\
{x}_{i}[2]  \\
\end{array}%
\!\!\right] \!\!+\!\! \left[\!\!%
\begin{array}{c}
{z}_{k}[1]   \\
{z}_{k}[2]  \\
{z}_{k}[3]  \\
{z}_{k}[4]  \\
{z}_{k}[5]  \\
\end{array}%
\!\!\right]. \label{eq:example_rx}
\end{align} 
By taking two output signals $y_k[3]+y_k[5]$ and $y_k[4]$, we obtain 
\begin{align}
\underbrace{\left[\!\!\!%
\begin{array}{c}
{y}_{k}[3]  \!+\! {y}_{k}[5]  \\
{y}_{k}[4]  \\
\end{array}%
\!\!\!\right] }_{ {\bf \bar y}_{k}}
\!&\!= \!\!\underbrace{\left[\!\!\!%
\begin{array}{cc}
{h}_{k,k}[3]    \!\!&\!\! {h}_{k,k}[2]  \\
0 \!\!&\!\! {h}_{k,k}[3] \\
\end{array}%
\!\!\!\right]}_{{\bf \bar H}_{k,k}} \!\underbrace{ \left[\!\!\!%
\begin{array}{c}
{x}_{k}[1]   \\
{x}_{k}[2]  \\
\end{array}%
\!\!\!\right]}_{{\bf x}_k} \nonumber \\\!&\!\!\!\!\!\!\!\!\!\!+ \sum_{\ell\neq k}^K\underbrace{\left[\!\!%
\begin{array}{cc}
{h}_{k,i}[1]  \!\!&\!\! {h}_{k,i}[2] \\
{h}_{k,i}[2]  \!\!&\!\! {h}_{k,i}[1] \\
\end{array}%
\!\!\right] }_{{\bf 
\bar H}_{k,i}} \!\underbrace{\left[\!\!\!%
\begin{array}{c}
{x}_{i}[1]   \\
{x}_{i}[2]  \\
\end{array}%
\!\!\!\right]}_{{\bf x}_{i}} \!+\! \underbrace{\left[\!\!\!\!%
\begin{array}{c}
{z}_{k}[3] \!+\!{z}_{k}[5]  \\
{z}_{k}[4]  \\
\end{array}%
\!\!\!\!\right]}_{{\bf \bar z}_k}, \nonumber
\end{align}
which reveals that the channel matrices for the interfering links can be made circulant, while the matrix for the desired link is kept non-circulant. By applying the same decoding strategy as explained in Example 1, the sum-DoF of $\frac{K}{5}$ is achievable.  

As shown in Example 2, one can still use the proposed idea in the case of $L_{\rm D} < L_{\rm I}$, provided that the support sets $\{ {h}_{k,k}[1], h_{k,k}[2],\ldots, h_{k,k}[L_{\rm D}]\}$ and $ \left\{ {h}_{k,i}[1], h_{k,i}[2],\ldots, h_{k,i}[L_{{\rm I}}] \right\}$ differs, i.e., $|{\rm supp} \left( {h}_{k,i}[1], h_{k,i}[2],\ldots, h_{k,i}[L_{{\rm I}}]\right) |  < |{\rm supp} \left( {h}_{k,k}[1], h_{k,k}[2],\ldots, h_{k,k}[L_{{\rm d}}]\right) | $.

In the rest of this paper, however, we will focus on the case of $L_{\rm D} > L_{\rm I}$ by assuming no relative propagation delay occurs between the desired and the interfering links. This assumption implies that all channel coefficients have non-zero entries, i.e., $h_{k,k}[\ell]\neq 0$ for $\ell\in\{1,\ldots,L_{k,k}\}$ and $h_{k,i}[\ell]\neq 0$ for $\ell\in\{1,\ldots,L_{k,i}\}$. This assumption made for mathematical convenience to explain the proposed idea clearly, but does not preclude application of the proposed idea in the case of $L_{\rm I} > L_{\rm D}$ with relative propagation delay.

% This condition, in fact, can be relaxed in a few ways and it is feasible to apply the proposed idea even in the case of $L_{\rm I} > L_{\rm D}$.

%One major limitation of the proposed IF-OFDM transmission applied to wireless networks may be the requirement that the number of channel-taps (or delay spreads) of desired link is lager than that of interfering links. 

%One possible relaxed condition is that the number of non-zero support elements of $\{ {h}_{k,k}[1], h_{k,k}[2],\ldots, h_{k,k}[L_{\rm D}]\}$ is larger than that of $ \left\{ {h}_{k,i}[1], h_{k,i}[2],\ldots, h_{k,i}[L_{{\rm I}}] \right\}$, i.e., $|{\rm supp} \left( {h}_{k,i}[1], h_{k,i}[2],\ldots, h_{k,i}[L_{{\rm I}}]\right) |  < |{\rm supp} \left( {h}_{k,k}[1], h_{k,k}[2],\ldots, h_{k,k}[L_{{\rm d}}]\right) | $, where ${\rm supp}({\bf x})$ denotes a support index set of ${\bf x}$. We present how the proposed method is applicable under this relaxed condition by providing an example. 
 
\subsection{Remarks}

 {\bf Remark 1 (No CSIT and No CSIR for interfering links):} The most prominent feature of IF-OFDM is that all IUI signals are canceled with the common and the predefined transmit beamforming vector ${\bf f}_1$  and receive beamforming vector ${\bf f}_2$. This feature implies that all transmitters and receivers do not require to have any knowledge in both either CSIT and CSIR to the designing of ${\bf f}_1$ and ${\bf f}_2$. This fact makes the proposed idea extremely practical.

%Therefore, the proposed idea is discernible compared to the  conventional blind interference alignment schemes [x,x] in which essentially needs special assumptions on channel coherence structure in both time and frequency domains [] or needs additional hardware for performing antenna switching [].

 {\bf Remark 2 (Interpretation of the proposed method with the lens of multi-carrier systems):} As illustrated in Example 1 and 2, all transmitters send a data symbol along with the first column vector of the 2-point IDFT matrix ${\bf f}_1=[1,~1]^{\top}$. This beamforming method can be interpreted as the spreading operation of a data symbol (in the first subcarrier) to a time-domain signal vector with length two; this is similar to a conventional multicarrier transmission method. Because ${\bf f}_1$ is the first eigenvector of ${\bf \bar H}_{k,i}$, all the interference signals at receiver $k\in\mathcal{K}$ are aligned to its direction; this alignment implies that they remain in the first subcarrier. Meanwhile, the non-circulant channel matrix ${\bf \bar H}_{k,k}$ spreads the desired signal (sent through the first subcarrier) to the both subcarriers. Receiver $k$ discards the aligned interference signal in the first-subcarrier by multiplying ${\bf f}_2^H$, which is the second eigenvector of ${\bf \bar H}_{k,i}$. Accordingly, the desired signal contained in the second subcarrier can be obtained.

%%%%%%%%%%%%%%%%%%%%%%%%%%%%%%%%%%%%%%%%%%%%%%%%%%%%%%%%%%%%%%%%%%%%%%%%%%%%%%%%%%%%%%%%%%%%%%%%%%%%%%%%%%%%%%%%%%%%%%%%%%%%%%%%%%%%%%%

\section{Main Result}
 
In this section, we generalize the idea of IF-OFDM introduced in the previous section. The following theorem is the main result of this paper.

 \begin{theorem} \label{Theorem1}
Consider a $K$-user wideband SISO interference channel with $L_{k,i}$. Let $L_{\rm D}=\max_k\{L_{k,k}\}$ and $L_{\rm I}=\max_k\max_i\{L_{k,i}\}$ for $k\neq i$, respectively. Then, the achievable sum-DoF with completely no CSIT is
\begin{align}
d^{\rm IC}_{\Sigma} = \max\left\{ \sum_{k\in\mathcal{K}}\frac{(L_{k,k}-L_{\rm I})^{+}}{(N+L_{\rm I}-1)}, 1 \right\},\label{eq:theorem1}
\end{align} 
where $N=\max\{L_{\rm I}, 2(L_{\rm D} -L_{\rm I}) \}$ and $(x)^+ = \max\{x, 0\}$.
\end{theorem}

We start by providing a Lemma that is essential for demonstrating our achievability result. The following lemma yields the matrix factorization of a circulant matrix. 

\begin{lemma}\label{lemma1} Let ${\bf C}\in\mathbb{C}^{n\times n}$ be a circulant matrix comprised of elements $\{c_0,c_2,\ldots, c_{n-1}\}$. Then, ${\bf C}$ is decomposed as
\begin{align}
{\bf C}= {\bf F}{\bf \Lambda}{\bf F}^H
\end{align}
where ${\bf F}=[{\bf f}_1,{\bf f}_2,\ldots,{\bf f}_n]\in\mathbb{C}^{n\times n}$ is the $n$-point IDFT matrix whose $k$th column vector is defined as
\begin{align}
{\bf f}_{k}=\frac{1}{\sqrt {n}}\left[1,~\omega _{k},~\omega _{k}^{2},~\ldots ,~\omega _{k}^{n-1}\right]^{\top},
\end{align} 
$k=\{0,1,\ldots ,n-1\}$ and  $\omega _{k}=\exp \left({-\frac {2\pi kj}{n}}\right)$ are the $n$-th roots of unity. The eigenvalue associated with ${\bf f}_k$ is then given by
\begin{align}
  \lambda _{k}=c_{0}+c_{n-1}\omega _{k}+c_{n-2}\omega _{k}^{2}+\ldots +c_{n}\omega _{k}^{n-1}.
\end{align}
\end{lemma}
\proof See \cite{Golub:96}.
\endproof

This lemma is used in the proof of Theorem \ref{Theorem1}.
\proof

In this proof, we focus on the case of $L_{k,k} > L_{\rm I}$ for some $k\in\mathcal{K}$ because it is trivial to achieving the sum-DoF$=1$ by TDMA-OFDM when $ \sum_{k\in\mathcal{K}}\frac{(L_{k,k}-L_{\rm I})^{+}}{(N+L_{\rm I}-1)} <1$.

The proposed achievable method uses a block transmission technique. Let $M$ be the total number of time slots per  transmission block. The transmission block with $M$ time slots consists of $B$ sub-blocks (or $B$ OFDM symbols) each with ${\bar N}=N +L_{\rm I}-1$ time slots, where $N=\max\left\{2(L_{\rm D}-L_{\rm I}),L_{\rm I}\right\}$. At the end of the transmission block, we span $L_{\rm D}-1$ additional time slots to prevent from inter-block interference. Therefore, the total number of time slots needed for a block transmission is
\begin{align}
M=   B{\bar N} + L_{\rm D}-1.
\end{align}
In this proof, we first demonstrate how each communication link achieves DoF of $d_{k}=\frac{(L_{k,k}-L_{{\rm I}})B}{B(N +L_{\rm I}-1)+ L_{\rm D}-1}$ when the number of subblocks $B$ is fixed. We present a  transmission method that uses an OFDM technique based on cyclic prefixes, because it is a more systematic transmission method than the one explained in Section III.

Let ${\bf \bar x}^b_k$ be the transmitted signal vector of the $b$th sub-block with length $N=\max\{ 2(L_{\rm D}-L_{\rm I}), L_{\rm I}\}$, which is defined as
\begin{align}
{\bf \bar x}^b_k\!=\! \left[\!\!
\begin{array}{cccc}
x_k[(b-1){\bar N}+1], 
\!\!&\!\!\cdots, \!\!&\!\! x_k[(b-1){\bar N}\!+\!N] \\ 
\end{array}%
\!\!\right]^{\!\!\top},
\end{align}
where $b\in\{1,2,\ldots,B\}$. We also use $\mathcal{S}_k=\{1,2,\ldots, L_{k,k}-L_{\rm I} \}$ to denote a set of subcarrier indices for delivering data symbols of transmitter $k$. We create ${\bf \bar x}^b_k$ by the superposition of $\left\{s_{k,1}^b,s_{k,2}^b,\ldots, s_{k,L_{k,k}-L_{\rm I}}^b\right\}$ information symbols along with precoding vectors $\left\{ {\bf f}_{1},{\bf f}_2,\ldots, {\bf f}_{L_{k,k}-L_{\rm I}} \right\}$, which is
\begin{align}
{\bf \bar x}_k^b= \sum_{n=1}^{L_{k,k}-L_{\rm I}}{\bf f}_{n}s^b_{k,n},
\end{align}
where ${\bf f}_{n}=\frac{1}{\sqrt {N}}\left[1,~e^{{\frac {2\pi n}{N}}},~e^{{\frac {2\pi 2n}{N}}},~\ldots ,~e^{{\frac {2\pi n(N-1)}{N}}}\right]^{\top}$ is the $n$th eigenvector of a circulant matrix with size of an $N$ by $N$ as shown in Lemma \ref{lemma1}. One possible interpretation of this precoding method is that it spreads $L_{k,k}-L_{\rm I}$ information symbols in the frequency domain using a set of column vectors of the $N$-point IDFT matrix to multiplex the signals in the time domain.

We generate the input data vector of the $b$th sub-block with length ${\bar N}=N+L_{{\rm I}}-1$ by adding a cyclic prefix with length $L_{{\rm I}}-1$ to ${\bf \bar x}_k^{b}$, i.e.,
 \begin{align}
{\bf x}_k^b\!=\!\left[{\bf \bar x}_k^{b,{\rm cp}}, ~ {\bf \bar x}_k^{b} \right]^{\!\top},
\end{align}
where ${\bf \bar x}_k^{b,{\rm cp}}\!\!=\!\!
\left[x_k[(b\!-\!1\!){\bar N}\!+\!N\!-\!L_{\rm I}\!+\!2],\ldots, x_k[(b\!-\!1){\bar N}\!+\!N]\right]^{\!\!\top}\!\!$. To prevent from inter-block-interference of the desired link, we provide a guard interval with length $L_{\rm D}-1$ zeroes. Finally, the transmitted signal vector from transmitter k during a block transmission is 
Finally, the transmitted signal vector from transmitter $k$ during a block transmission is 
 \begin{align}
{\bf x}_k\!=\!\left[\left({\bf  x}_k^{1}\right)^{\!\top},\left({\bf  x}_k^{2}\right)^{\!\top},\ldots,\left({\bf  x}_k^{B}\right)^{\!\top},\underbrace{0,\ldots, 0}_{L_{\rm D}-1} \right]^{\!\top}\!\!\!. 
\end{align}

Then, the received signal of receiver $k$ in time slot $n$ is expressed by the superposition of $K$ linear convolution operations between the transmitted signals and the channel impulse responses. Without loss of generality, we focus on the output signals of the $b$th sub-block at receiver $k$, which are
\begin{align}
y_{k}[(b\!-\!1){\bar N} + n] &\!= \sum_{\ell=1}^{L_{k,k}}h_{k,k}[\ell]x_{k}[(b\!-\!1){\bar N} + n-\ell+1] \nonumber \\
\!&\!+\!\!\sum_{i \in\mathcal{K}/\{k\} }  \sum_{\ell=1}^{L_{k,i}}h_{k,i}[\ell]x_{i}[(b\!-\!1){\bar N}\!+\!n\!-\!\ell+1]  \nonumber \\
\!&\!+ z_k[(b\!-\!1){\bar N} +t], \nonumber 
\end{align}
where $n\in\{1,2,\ldots, {\bar N}\}$ and $b\in\{1,2,\ldots, B\}$. Let ${\bf \bar y}_k^b=\! \left[ 
y_k[(b\!-\!1)B\!+\!L_{\rm I}],\cdots,y_k[(b\!-\!1)B\!+\!{\bar N}] \right]^{\!\top}\in\mathbb{C}^N$ and ${\bf \bar z}_k^b=\! \left[
z_k[(b\!-\!1)B\!+\!L_{\rm I}], 
 \cdots,z_k[(b\!-\!1)B\!+\!{\bar N} 
\right]^{\!\!\top} \in\mathbb{C}^N$ be the effective received signal and noise vectors of the $b$th sub-block after discarding the cyclic prefix, respectively. Then, the effective input-output relationship of the $b$th sub-block in a matrix form is given by 
\begin{align}
{\bf \bar y}_k^b \!&\!= {\bf \bar H}^b_{k,k}{\bf \bar x}^b_k + \sum_{i=1,i\neq k}^K{\bf \bar H}^b_{k,i}{\bf \bar x}^b_i + {\bf \bar z}^b_k, \label{eq:cp_remove}
\end{align} 
where
\begin{align}\small
{\bf \bar H}_{k,i}^b\!\!=\!\!\!\left[\!\!\!\!%
\begin{array}{ccccccc}
{h}_{k,i}[1]  \!\!&\!\! \cdots  \!\!&\!\!  {h}_{k,i}[L_{k,i}]  \!\!&\!\!  \cdots \!\!&\!\!  {h}_{k,i}[3] \!\!&\!\!  {h}_{k,i}[2] \\
{h}_{k,i}[2]  \!\!&\!\!  {h}_{k,i}[1]  \!\!&\!\! \cdots \!\!&\!\! {h}_{k,i}[L_{k,i}] \!\!&\!\!  \cdots \!\!&\!\!   {h}_{k,i}[3]\\
\vdots  \!\!&\!\!   \vdots  \!\!&\!\!  \ddots\!\!&\!\!  \ddots   \!\!&\!\!   \ddots \!\!&\!\!   \vdots \\
{h}_{k,i}[L_{k,i}] \!\!&\!\!  0 \!\!&\!\!  \ddots \!\!&\!\!  \cdots \!\!&\!\! 0 \!\!&\!\!   {h}_{k,i}[L_{k,i}]\\
0 \!\!&\!\!   {h}_{k,i}[L_{k,i}] \!\!&\!\! 0 \!\!&\!\!  \cdots \!\!&\!\! {h}_{k,i}[1] \!\!&\!\!   0\\
0 \!\!&\!\!  0 \!\!&\!\!  {h}_{k,i}[L_{k,i}\!] \!\!&\!\!  \cdots \!\!&\!\! {h}_{k,i}[2] \!\!&\!\!   {h}_{k,i}[1]\\
\end{array}%
\!\!\!\!\right] \nonumber
\end{align} is the effective channel matrix from transmitter $i$ to receiver $k$ with size of $N\times N$ for all $i,k\in\mathcal{K}$ and $i\neq k$. It is important note that ${\bf \bar H}_{k,i}^b$ is a circulant matrix, because the length of ${\bf \bar x}_k^{b,{\rm cp}}$ was selected with length $L_{{\rm I}} -1 \geq \max_{k}\max_{i}\{L_{k,i}\}-1$. However, the effective channel matrix that carries the desired information symbols, ${\bf \bar H}^b_{k,k}\in\mathbb{C}^{N\times N}$ does not hold the circulant matrix structure, because the length of cyclic prefix vector ${\bf \bar x}_k^{b,{\rm cp}}$ was chosen to be shorter than the  length of channel-taps of direct links, i.e., $  L_{\rm I} <L_{k,k}$. This creates \textit{the relativity of alignment between direct channels and interference channels by a matrix structure.}

 \textbf{Inter-User Interference Cancellation:}

We have constructed the input data vector of the $b$th sub-block as the superposition of $L_{\rm k,k}-L_{\rm I}$ data streams by using linear beamforming, i.e., ${\bf \bar x}_k^b= \sum_{n=1}^{L_{k,k}-L_{\rm I}}{\bf f}_{n}s^b_{k,n}\in\mathbb{C}^{N}$ to perform the IF-OFDM transmission. Utilizing this, the received signal expression in \eqref{eq:cp_remove} is rewritten as
\begin{align}
{\bf \bar y}_k^b \!&\!= {\bf \bar H}_{k,k}^b\sum_{n=1}^{L_{k,k}-L_{\rm I}}\!\!{\bf f}_{n}s_{k,n}^b \!+\! \sum_{i=1,i\neq k}^K{\bf \bar H}_{k,i}^b\!\!\sum_{n=1}^{L_{k,k}-L_{\rm I}}{\bf f}_{n}s_{i,n}^b+ {\bf \bar z}_k^b,  \nonumber \\ 
\!&\!= {\bf \bar H}_{k,k}^b\sum_{n=1}^{L_{k,k}-L_{\rm I}}{\bf f}_{n}s_{k,n}^b +\sum_{i=1,i\neq k}^K\!\! \sum_{n=1}^{L_{k,k}-L_{\rm I}}\lambda_{k,i}^{n,b}\!\!{\bf f}_{n}s_{i,n}^b+ {\bf \bar z}_k^b, \label{eq:cp_remove2}
\end{align} 
where the second equality follows from the fact that ${\bf \bar H}_{k,i}^b{\bf f}_{n}=\lambda_{k,i}^{n,b}{\bf f}_{n}$ for all $i,k\in \mathcal{K}$, $i\neq k$, and $n\in\{1,2,\ldots,L_{k,k}-L_{\rm I}\}$. Notice that the interference signals sent from transmitter $i$ are confined at receiver $k\in\mathcal{K}$ in an $(L_{k,k}-L_{\rm I})$-dimensional subspace defined as
\begin{align}
\mathcal{I}_{k,i} =: {\rm Span}\left(
\left[
\begin{array}{cccc}
{\bf f}_{1} \!&\!  {\bf f}_{2} \!&\! \cdots \!&\!{\bf f}_{L_{k,k}-L_{\rm I}} 
\end{array}%
\!\!\!\right]\right).
\end{align}
Because the union of $\mathcal{I}_{k,i}$ belongs to the total interference subspace, namely
\begin{align}
\mathcal{I}_{k} =: \cup_{i\neq k}^K\mathcal{I}_{k,i}  \subseteq {\rm Span}\left(
\left[
\begin{array}{cccc}
{\bf f}_{1} \!&\!  {\bf f}_{2} \!&\! \cdots \!&\!{\bf f}_{L_{{\rm D}}-L_{\rm I}} 
\end{array}%
\!\!\!\right]\right),
\end{align}
we eliminate all IUI signals sent from $K-1$ transmitters at receiver $k$ by multiplying ${\bf F}_{\mathcal{S}_k^c}^H=\left[{\bf f}_{L_{\rm D}-L_{\rm I}+1},~ \cdots,~{\bf f}_{L_{k,k}+L_{\rm D}-2L_{\rm I})} \right]^H\in\mathbb{C}^{(L_{k,k}-L_{\rm I}) \times N} $, which spans the null space of $\mathcal{I}_k$, to ${\bf \bar y}_k$; thereby, the output signal vector that only contains the desired data symbols ${\bf s}^b_{k}=[s_{k,1}^b,~s_{k,2}^b,\ldots,s_{k,L_{\rm D}-L_{\rm I}}^b]^{\top}$ can be obtained as
\begin{align}
{\bf \tilde y}_k^b \!&\!= {\bf F}_{\mathcal{S}_k^c}^H{\bf \bar y}_k^b \nonumber \\ \!&\!= {\bf F}_{\mathcal{S}_k^c}^H{\bf \bar H}_{k,k}^b{\bf F}_{\mathcal{S}_k}{\bf s}_{k}^b +{\bf \tilde z}_k^b,\label{eq:cp_remove3}
\end{align} 
where ${\bf F}_{\mathcal{S}_k}=[{\bf f}_{1},~ {\bf f}_{2},~ \cdots,~{\bf f}_{L_{k,k}-L_{\rm I}} ]\in\mathbb{C}^{N\times (L_{k,k}-L_{\rm I})}$ and ${\bf \tilde z}_k^b={\bf F}_{\mathcal{S}_k^c}^H {\bf \bar z}_k^b$ is the effective noise vector at receiver $k$ whose distribution is invariant with ${\bf \bar z}_k^b$ because ${\bf F}_{\mathcal{S}_k^c}^H$ is a unitary transformation matrix.

\textbf{Decodability of Subblock Data:}
Now, the remaining key step to end this proof is to show that ${\rm rank}\left( {\bf F}_{\mathcal{S}_k^c}^H{\bf \bar H}_{k,k}^b{\bf F}_{\mathcal{S}_k}\right)=L_{k,k}  - L_{\rm I}$ almost surely. We prove this by considering two different cases: 1) $N \geq L_{\rm D}$ and 2) $L_{\rm I} \leq N < L_{\rm D}$.

\subsubsection{Case of $N \geq L_{\rm D}$}

We first consider the case of $N \geq L_{\rm D}$, i.e., $L_{\rm D}\geq 2L_{\rm I}$. In this case, by the definition, ${\bf \bar H}^b_{k,k}$ can be written as in \eqref{eq:direct_channel_matrix1}.
\begin{figure*}
\begin{equation}\small
{\bf \bar H}^b_{k,k}\!=\small\!\!\left[\!\!\!%
\begin{array}{ccccccccc}
{h}_{k,k}[1]  \!\!&\!\! 0  \!\!&\!\!  \cdots\!\!&\!\! \cdots &\!\! \cdots \!\!&\!\!  {h}_{k,k}[L_{\rm I}] \!\!&\!\!  \cdots \!\!&\!\!  {h}_{k,k}[2] \\
{h}_{k,k}[2]  \!\!&\!\!  {h}_{k,k}[1]  \!\!&\!\! 0 \!\!&\!\!  \cdots &\!\! \cdots \!\!&\!\!  {h}_{k,k}[L_{\rm I}-1] \!\!&\!\!  \cdots \!\!&\!\!   {h}_{k,k}[3]\\
{h}_{k,k}[3]  \!\!&\!\!  {h}_{k,k}[2]  \!\!&\!\! {h}_{k,k}[1] \!\!&\!\!  \ddots \!\!&\!\!  \vdots \!\!&\!\!  \cdots&\!\! \cdots \!\!&\!\!  \vdots \\
\vdots  \!\!&\!\!  \vdots \!\!&\!\! \ddots \!\!&\!\!  \ddots &\!\! \cdots \!\!&\!\!  h_{k,k}[L_{k,k}] \!\!&\!\!  \ddots \!\!&\!\!   \vdots\\
\vdots  \!\!&\!\!  \vdots \!\!&\!\! \ddots &\!\! \cdots \!\!&\!\!  \ddots \!\!&\!\!  0 \!\!&\!\!  \ddots \!\!&\!\!   {h}_{k,k}[L_{k,k}-1]\\
\vdots  \!\!&\!\!  \vdots \!\!&\!\! \ddots \!\!&\!\!  \ddots \!\!&\!\!  \ddots \!\!&\!\! \vdots &\!\!  \cdots \!\!&\!\!   {h}_{k,k}[L_{k,k}]\\
{h}_{k,k}[L_{k,k}]  \!\!&\!\!  {h}_{k,k}[L_{k,k}\!-\!1]  \!\!&\!\! \ddots \!\!&\!\!  \ddots \!\!&\!\!  \cdots &\!\! 0 \!\!&\!\!  \cdots \!\!&\!\!   0\\
0  \!\!&\!\!  {h}_{k,k}[L_{k,k}]  \!\!&\!\! \ddots \!\!&\!\!  \ddots \!\!&\!\!  \vdots&\!\! \vdots  \!\!&\!\!  \cdots \!\!&\!\!   0\\
\vdots  \!\!&\!\!   \vdots  \!\!&\!\!  \ddots\!\!&\!\!  \ddots   \!\!&\!\!   \ddots &\!\! 0 \!\!&\!\!   \vdots  \!\!&\!\!   \vdots \\
0 \!\!&\!\!  0 \!\!&\!\!  {h}_{k,k}[L_{k,k}] \!\!&\!\!   h_{k,k}[L_{k,k}\!-\!1] &\!\! \cdots \!\!&\!\!  \cdots \!\!&\!\! \cdots \!\!&\!\!   {h}_{k,k}[1]\\\hline
\end{array}%
\!\!\!\right]\in \mathbb{C}^{N\times N}. \vspace{-0.3cm}\label{eq:direct_channel_matrix1}
\end{equation}
\end{figure*} 
By the linearity of a linear-time-invariant system, ${\bf \bar H}^b_{k,k}$ in \eqref{eq:direct_channel_matrix1} can be decomposed into the difference of the circulant and the non-circulant matrices as
\begin{align}
{\bf \bar H}^b_{k,k} = {\bf \bar H}^{b,{\rm C}}_{k,k} - {\bf \bar H}^{b,{\rm NC}}_{k,k},
\end{align} 
where 
\begin{align}
{\bf \bar H}^{b,{\rm C}}_{k,k} = {\bf I}_{N} h_{k,k}[1] + {\bf P}^1h_{k,k}[2] +\cdots + {\bf P}^{L_{\rm D}}h_{k,k}[L_{k,k}] \nonumber
\end{align}
and 
${\bf P}_{N}^i$ is the $i$th power of the $N \times N$ cyclic permutation matrix 
\begin{align}
{\bf P}=\!\left[\!\!\!%
\begin{array}{ccccc}
0  & 0    & \cdots &  0 &  1 \\
1  & 0    & \cdots &  0 &  0 \\
0  & 1    & \cdots &  0 &  0 \\
\vdots  & \vdots    & \ddots &  \vdots &  \vdots  \\
0  & 0    & \cdots &  1 &  0 \\
\end{array}%
\!\!\!\right].  \nonumber
\end{align}
Furthermore, the non-circulant matrix ${\bf \bar H}^{b,{\rm NC}}_{k,k}\in\mathbb{C}^{N\times N}$ is defined as
\begin{align}
{\bf \bar H}^{b,{\rm NC}}_{k,k}=\!\left[\!\!\!%
\begin{array}{ccc}
{\bf 0}_{\frac{N}{2} \times (N-L_{k,k}+1)} & {\bf \hat H}^{b,{\rm NC}}_{k,k}      &  {\bf 0}_{\frac{N}{2} \times (L_{\rm I}-1)} \\
{\bf 0}_{\frac{N}{2} \times (N-L_{k,k}+1)}  & {\bf 0}_{\frac{N}{2} \times (L_{k,k}-L_{\rm I})}      &  {\bf 0}_{\frac{N}{2} \times (L_{\rm I}-1)} \\
\end{array}%
\!\!\!\right],  \nonumber
\end{align}
where the submatrix ${\bf \hat H}^{b,{\rm NC}}_{k,k}\in\mathbb{C}^{\frac{N}{2}\times \frac{N}{2}}$ is an upper triangular matrix, namely,
\begin{align}\small
\small {\bf \hat H}^{b,{\rm NC}}_{k,k}\!=\!\left[\!%
\begin{array}{cccc}
h_{k,k}[L_{k,k}]  \!\!&\!\! h_{k,k}[L_{k,k}\!-\!1]   \!\!&\!\! \cdots  \!\!&\!\!  h_{k,k}[ L_{\rm I}+1]  \\
0  \!\!&\!\! h_{k,k}[L_{k,k}]   \!\!&\!\! \cdots  \!\!&\!\!  h_{k,k}[L_{\rm I}+2]  \\
0  \!\!&\!\! 0  \!\!&\!\! \cdots  \!\!&\!\!  h_{k,k}[L_{\rm I}+3]  \\
\vdots  \!\!&\!\! \vdots   \!\!&\!\! \ddots\!\!&\!\!   \vdots  \\
0 \!\!&\!\!0   & \cdots  \!\!&\!\!  h_{k,k}[L_{k,k}]  \\
\end{array}%
\!\!\right].  \label{eq:non_circulant_1} 
\end{align}
From this decomposition, it implies that
\begin{align}
 {\bf F}_{\mathcal{S}_k^c}^H{\bf \bar H}_{k,k}^b{\bf F}_{\mathcal{S}_k} &= {\bf F}_{\mathcal{S}_k^c}^H\left( {\bf \bar H}^{b,{\rm C}}_{k,k} - {\bf \bar H}^{b,{\rm NC}}_{k,k}\right) {\bf F}_{\mathcal{S}_k}  \nonumber \\
&= -{\bf F}_{\mathcal{S}_k^c}^H {\bf \bar H}^{b,{\rm NC}}_{k,k} {\bf F}_{\mathcal{S}_k}. 
\end{align}
Notice that $h_{k,k}[L_{\rm I}+1],h_{k,k}[L_{\rm I}+2],\ldots, h_{k,k}[L_{k,k}]$ in \eqref{eq:non_circulant_1} were selected from IID complex random variables, i.e., ${\rm rank}\left( {\bf \bar H}^{b,{\rm NC}}_{k,k}\right)=L_{k,k}-L_{\rm I}$. Furthermore, because ${\bf F}_{\mathcal{S}_k}$ and ${\bf F}_{\mathcal{S}_k^c}^H$ are the sub-matrices of the $N$-point IDFT and DFT matrix, then with high probability, the rank of the resultant channel matrix is
\begin{align}
{\rm rank}\left( {\bf F}_{\mathcal{S}_k^c}^H {\bf \bar H}^{b}_{k,k} {\bf F}_{\mathcal{S}_k}\right) &={\rm rank}\left( -{\bf F}_{\mathcal{S}_k^c}^H {\bf \bar H}^{b,{\rm NC}}_{k,k} {\bf F}_{\mathcal{S}_k}\right)\nonumber \\&= L_{k,k}-L_{\rm I}.
\end{align}

%%%%%%%%%%%%%%%%%%%%%%%%%
\subsubsection{Case of $L_{\rm I} \leq N < L_{\rm D}$}

In this case, after discarding the cyclic prefix with size $L_{\rm I}-1$, the effective channel matrix between transmitter $k$ and receiver $k$ can be represented as the sum of two matrices, namely,
\begin{align}
{\bf \bar H}^b_{k,k} = {\bf \bar H}^{b,{\rm data}}_{k,k} +{\bf \bar H}^{b,{\rm cp}}_{k,k},
\end{align}
where 
\begin{align}\small
{\bf \bar H}^{b,{\rm data}}\!\!=\!\!\left[\!\!\!\!%
\begin{array}{cccccc}
{h}_{k,k}[1]  \!\!&\!\! 0  \!\!&\!\! \cdots \!\!&\!\! 0  \!\!&\!\! 0  \!\!&\!\! 0  \\
{h}_{k,k}[2]  \!\!&\!\! {h}_{k,k}[1]  \!\!&\!\! 0   \!\!&\!\! 0  \!\!&\!\! \cdots   \!\!&\!\! 0 \\
\vdots \!\!&\!\! \ddots  \!\!&\!\!  \ddots \!\!&\!\! \vdots \!\!&\!\! \vdots  \!\!&\!\! \vdots \\
\vdots \!\!&\!\! \ddots  \!\!&\!\!  \ddots\!\!&\!\! \ddots  \!\!&\!\! 0   \!\!&\!\! \vdots \\
{h}_{k,k}[N\!\!-\!\!1]    \!\!&\!\! \ddots\!\!&\!\! \ddots   \!\!&\!\! {h}_{k,k}[2] \!\!&\!\!  {h}_{k,k}[1]  \!\!&\!\! 0  \\
 {h}_{k,k}[N] \!\!&\!\! {h}_{k,k}[N\!\!-\!\!1]   \!\!&\!\!  \cdots \!\!&\!\! \cdots \!\!&\!\!  h_{k,k}[2]  \!\!&\!\!  {h}_{k,k}[1]  \\
\end{array}%
\!\!\!\!\right], \nonumber 
\end{align}
denotes the data matrix, and 
\begin{align}\small
{\bf \bar H}^{b,{\rm cp}}_{k,k}\!\!=\!\!\left[\!\!\!\!%
\begin{array}{cccccccc}
0 \!\!&\!\! \cdots  \!\!&\!\!   0  \!\!&\!\!  {h}_{k,k}[\!L_{\rm I}\!]  \!\!\!&\!\!\! {h}_{k,k}[\!L_{\rm I}\!\!-\!\!1\!]    \!\!&\!\! \cdots   \!\!\!&\!\!\! h_{k,k}[3]   \!\!\!&\!\!\!h_{k,k}[2]  \\
0  \!\!&\!\!  \cdots  \!\!&\!\!  0  \!\!&\!\!  \vdots \!\!&\!\! \ddots  \!\!\!&\!\!\!  \ddots \!\!\!&\!\!\!h_{k,k}[4] \!\!\!&\!\!\! h_{k,k}[3]  \\
0  \!\!&\!\!  \cdots \!\!&\!\!  0  \!\!&\!\!   {h}_{k,k}[\!L_{k,k}\!\!-\!\!1\!]  \!\!\!&\!\!\! \ddots   \!\!\!&\!\!\! \vdots \!\!\!&\!\!\! \ddots   \!\!\!&\!\!\!  h_{k,k}[4]  \\
0  \!\!&\!\!  \cdots  \!\!&\!\!  0  \!\!&\!\!   {h}_{k,k}[L_{k,k}]    \!\!\!&\!\!\! \vdots \!\!\!&\!\!\!\ddots \!\!\!&\!\!\!  \vdots \!\!\!&\!\!\! \vdots  \\
0  \!\!&\!\!  \cdots  \!\!&\!\!  0  \!\!&\!\!  \vdots \!\!\!&\!\!\! \ddots  \!\!\!&\!\!\!  \vdots  \!\!\!&\!\!\! {h}_{k,k}[\!N\!\!+\!\!1\!]   \!\!\!&\!\!\! {h}_{k,k}[\!N\!]  \\
0  \!\!&\!\!  \cdots  \!\!&\!\!  0  \!\!&\!\!   0\!\!\!&\!\!\!  \cdots  \!\!\!&\!\!\!{h}_{k,k}[\!L_{k,k}\!]   \!\!\!&\!\!\! \cdots  \!\!\!&\!\!\! {h}_{k,k}[\!N\!\!+\!\!1\!]   \\
\end{array}%
\!\!\!\!\right]\!\! \nonumber 
\end{align}
denotes the cyclic prefix addition matrix. Similar to the previous case, we decompose this effective channel matrix into the sum of a circulant and a non-circulant matrix as
\begin{align}
{\bf \bar H}^b_{k,k}&= {\bf \bar H}^{b,{\rm data}}_{k,k} +{\bf \bar H}^{b,{\rm cp}}_{k,k} \nonumber\\
 &= {\bf \bar H}^{b,{\rm C}_2}_{k,k} + {\bf \bar H}^{b,{\rm NC}_2}_{k,k},
\end{align} 
where 
\begin{align}
{\bf \bar H}^{b,{\rm C}_2}_{k,k} = {\bf I}_{N} h_{k,k}[1] + {\bf P}^1h_{k,k}[2] +\cdots + {\bf P}^{N-1}h_{k,k}[N]. \nonumber
\end{align}
The non-circulant matrix is given as in \eqref{eq:NC_2}.
\begin{figure*}
\begin{align}\small
{\bf \bar H}^{b,{\rm NC}_2}_{k,k}=\left[\!\!%
\begin{array}{cccccccccc}
0  \!\!&\!\!  -{h}_{k,k}[N]    \!\!&\!\! \cdots  \!\!&\!\!  \cdots \!\!&\!\!   -{h}_{k,k}[\!L_{\rm I}\!+\!1\!]   \!\!&\!\!  0 \!\!\!&\!\!\! 0   \!\!&\!\! \cdots   \!\!\!&\!\!\! \cdots   \!\!\!&\!\!\!0  \\
0  \!\!&\!\!  0    \!\!&\!\! -{h}_{k,k}[N] \!\!&\!\!  \cdots \!\!&\!\!   -{h}_{k,k}[\!L_{\rm I}]   \!\!&\!\!  0 \!\!\!&\!\!\! 0   \!\!&\!\! \cdots   \!\!\!&\!\!\! \cdots   \!\!\!&\!\!\!0  \\
0  \!\!&\!\! 0  \!\!&\!\!    \cdots  \!\!&\!\!  \ddots \!\!&\!\!  \vdots \!\!&\!\!  \vdots \!\!&\!\! \ddots  \!\!\!&\!\!\!  \ddots \!\!\!&\!\!\!0\!\!\!&\!\!\! 0 \\
0  \!\!&\!\! 0  \!\!&\!\!    \cdots  \!\!&\!\!  0 \!\!&\!\!  -h_{k,k}[N] \!\!&\!\!  \vdots \!\!&\!\! \ddots  \!\!\!&\!\!\!  \ddots \!\!\!&\!\!\!0\!\!\!&\!\!\! 0 \\
0  \!\!&\!\! 0  \!\!&\!\!    \cdots  \!\!&\!\!  0 \!\!&\!\!  0 \!\!&\!\!  0\!\!&\!\! \cdots  \!\!\!&\!\!\!  \ddots \!\!\!&\!\!\!0\!\!\!&\!\!\! 0 \\
0  \!\!&\!\!0  \!\!&\!\!  0 \!\!&\!\!  \cdots \!\!&\!\!  0  \!\!&\!\!   {h}_{k,k}[\!N\!\!+\!\!1\!]  \!\!\!&\!\!\! \ddots   \!\!\!&\!\!\! \vdots \!\!\!&\!\!\! \ddots   \!\!\!&\!\!\!  0  \\
0  \!\!&\!\!0  \!\!&\!\!  0 \!\!&\!\!  \cdots \!\!&\!\!  0  \!\!&\!\!   \vdots  \!\!\!&\!\!\! \ddots   \!\!\!&\!\!\! \vdots \!\!\!&\!\!\! \ddots   \!\!\!&\!\!\!  0  \\
0  \!\!&\!\!0 \!\!&\!\!  0 \!\!&\!\!  \cdots  \!\!&\!\!  0  \!\!&\!\!   {h}_{k,k}[L_{k,k}]    \!\!\!&\!\!\! \vdots \!\!\!&\!\!\!\ddots \!\!\!&\!\!\!  \vdots \!\!\!&\!\!\! \vdots  \\
0  \!\!&\!\!0  \!\!&\!\!  0 \!\!&\!\!  \cdots  \!\!&\!\!  0  \!\!&\!\!  \vdots \!\!\!&\!\!\! \ddots  \!\!\!&\!\!\!  \vdots  \!\!\!&\!\!\! {h}_{k,k}[\!N\!\!+\!\!1\!]   \!\!\!&\!\!\! 0 \\
0  \!\!&\!\!0  \!\!&\!\!  0 \!\!&\!\!  \cdots  \!\!&\!\!  0  \!\!&\!\!   0\!\!\!&\!\!\!  \cdots  \!\!\!&\!\!\!{h}_{k,k}[\!L_{k,k}\!]   \!\!\!&\!\!\! \cdots  \!\!\!&\!\!\! {h}_{k,k}[\!N\!\!+\!\!1\!]   \\
\end{array}%
\!\!\right]\!\!. \label{eq:NC_2}  
\end{align}\vspace{-0.3cm}
\end{figure*}  
From this decomposition, the resultant channel matrix of the $k$th link is
\begin{align}
 {\bf F}_{\mathcal{S}_k^c}^H{\bf \bar H}_{k,k}^b{\bf F}_{\mathcal{S}_k} &= {\bf F}_{\mathcal{S}_k^c}^H\left( {\bf \bar H}^{b,{\rm C}_2}_{k,k} + {\bf \bar H}^{b,{\rm NC}_2}_{k,k}\right) {\bf F}_{\mathcal{S}_k}  \nonumber \\
&= {\bf F}_{\mathcal{S}_k^c}^H {\bf \bar H}^{b,{\rm NC}_2}_{k,k} {\bf F}_{\mathcal{S}_k}. 
\end{align}
Since $h_{k,k}[L_{\rm I}+1],h_{k,k}[L_{\rm I}+2],\ldots, h_{k,k}[L_{k,k}]$ were chosen to be IID complex random variables, ${\rm rank}\left( {\bf \bar H}^{b,{\rm NC}_2}_{k,k}\right)=L_{k,k}-L_{\rm I}$. Accordingly, we conclude that 
\begin{align}
{\rm rank}\left( {\bf F}_{\mathcal{S}_k^c}^H {\bf \bar H}^{b}_{k,k} {\bf F}_{\mathcal{S}_k}\right)&={\rm rank}\left( {\bf F}_{\mathcal{S}_k^c}^H {\bf \bar H}^{b,{\rm NC}_2}_{k,k} {\bf F}_{\mathcal{S}_k}\right)  \nonumber \\
&=L_{k,k}-L_{\rm I}.
\end{align}

\textbf{Inter-Subblock Interference Cancellation:}
We have shown that $L_{k,k}-L_{\rm I}$ independent data symbols in the $b$th sub-block are decodable after canceling the inter-user interference without CSIT, assuming that inter-subblock interference does not occur. Unfortunately, the inter-subblock interference is unavoidable between the $b$th subblock and its previous $(b\!-\!1)$th subblock for $b\in\{2,\ldots, B\}$ because the cyclic prefix size is shorter than the number of channel-taps of the direct link, i.e., $L_{\rm I}<L_{k,k}$.

By concatenating ${\bf \bar y}_k^b$ for $b\in\{1,2,\ldots, B\}$ after discarding $L_{\rm D}-1$ zeros at the end of the transmission block, the total input-output relationship during a transmission block when ignoring noise is  
\begin{align}
\left[\!\!\!%
\begin{array}{c}
{\bf \tilde y}_k^1 \\
{\bf \tilde y}_k^2 \\
\vdots \\
{\bf \tilde y}_k^{B}  \\
\end{array}%
\!\!\!\right]\!\!&=\!\! \left[\!\!\!%
\begin{array}{cccc}\small
 {\bf F}_{\mathcal{S}_k^c}^H{\bf \bar H}^1_{k,k}{\bf F}_{\mathcal{S}_k}\!&\! {\bf 0} \!&\!\cdots \!&\!  {\bf 0} \\
 {\bf F}_{\mathcal{S}_k^c}^H{\bf \bar H}^{1,2}_{k,k}{\bf F}_{\mathcal{S}_k}\!&\!{\bf F}_{\mathcal{S}_k^c}^H{\bf \bar H}^2_{k,k}{\bf F}_{\mathcal{S}_k}\!&\! \cdots  \!&\!  \!{\bf 0} \\
 {\bf 0}\!&\!{\bf F}_{\mathcal{S}_k^c}^H{\bf \bar H}^{3,2}_{k,k}{\bf F}_{\mathcal{S}_k}\!&\! \cdots \!&\!  {\bf 0} \\
 \vdots \!&\! \ddots \!&\! \ddots \!&\! \vdots \\
{\bf 0}\!&\! {\bf 0}\!&\! \cdots \!&\!    {\bf F}_{\mathcal{S}_k^c}^H{\bf \bar H}^{B}_{k,k}{\bf F}_{\mathcal{S}_k}\\
\end{array}%
\!\!\!\!\!\right]\!\!\!\!\left[\!\!\!\!%
\begin{array}{c}
{\bf s}_k^1 \\
{\bf s}_k^2 \\
\vdots \\
{\bf s}_k^{B}  \\
\end{array}%
\!\!\!\!\right]\!\!, \label{eq:concentrating_rx_signals}%\! \nonumber\\&
%+\!\left[\!\!%
%\begin{array}{c}
%{\bf \tilde z}^1_k \\
%{\bf \tilde z}^2_k \\
%\vdots \\
%{\bf \tilde z}^{B}_k  \\
%\end{array}%
%\!\right]\!\!,
\end{align}
where ${\bf H}_{k,k}^{b-1,b}\in\mathbb{C}^{N\times N}$ is the inter-subblock interference channel matrix of the $k$th link. By the definition, one can easily verify that it is the same matrix as in \eqref{eq:non_circulant_1}, 
\begin{align}
{\bf H}_{k,k}^{b-1,b}=\!\left[\!\!\!%
\begin{array}{cc}
{\bf 0}_{\frac{N}{2} \times (N-L_{k,k} + L_{\rm I})} & {\bf \hat H}^{b,{\rm NC}}_{k,k}   \\
{\bf 0}_{\frac{N}{2} \times (N-L_{k,k}+L_{\rm I})}  & {\bf 0}_{\frac{N}{2} \times (L_{k,k}-L_{\rm I})}    \\
\end{array}%
\!\!\!\right].
\end{align} 
Because no inter-subblock interference occurs in the the first subblock output vector, i.e., ${\bf \tilde y}_k^1= {\bf F}_{\mathcal{S}_k^c}^H{\bf \bar H}^1_{k,k}{\bf F}_{\mathcal{S}_k}{\bf s}_k^1+{\bf \tilde z}^1_k$ and ${\rm rank}\left({\bf F}_{\mathcal{S}_k^c}^H{\bf \bar H}^1_{k,k}{\bf F}_{\mathcal{S}_k}\right)=L_{k,k}-L_{\rm I}$, ${\bf s}_k^1$ can be reliably decoded when ignoring noise ${\bf \tilde z}^1_k$. Once ${\bf s}_k^1$ is decoded, ${\bf s}_k^2$ can be decoded using ${\bf \tilde y}_k^2 - {\bf F}_{\mathcal{S}_k^c}^H{\bf \bar H}^{2,1}_{k,k}{\bf F}_{\mathcal{S}_k}{\bf s}_k^1$ by successive interference cancellation. By applying this strategy over $B$ subblocks recursively, receiver $k$ is capable of decoding $B(L_{k,k}-L_{\rm I})$ independent data symbols, ${\bf s}_k^1, {\bf s}_k^2, \ldots, {\bf s}_k^{B}$, using $M=BN+L_{\rm D}-1$ channel uses. By symmetry, all the other receivers decode $B(L_{k,k}-L_{\rm I})$ independent data symbols spanning $M$ channel uses. As a result, as $B$ tends to infinity, we conclude that the achievable DoF of the $k$th communication link is 
\begin{align}
d_k&=\lim_{B\rightarrow \infty}\frac{B(L_{k,k}-L_{\rm I})}{(N+L_{\rm I}-1)B+L_{\rm D}-1} \nonumber\\
  &=\frac{L_{k,k}-L_{\rm I}}{(N+L_{\rm I}-1)}.
\end{align}
 Since $N=\max\left\{L_{\rm I}, 2(L_{\rm D}-L_{\rm I})\right\}$, we reach the expression in Theorem 1; this completes the proof.\endproof

We provide some remarks about our result.

 {\bf Remark 3 (Precoding matrix design and subcarrier selection):} The proposed transmission method suggests that all transmitters use the same precoding vectors when carrying their own data symbols. In our proof, for ease of exposition, we construct the precoding matrix by taking the first $L_{k,k}-L_{\rm I}$ column vectors among the first $L_{{\rm D}}-L_{\rm I}$ column vectors of a $N\times N$ IDFT matrix, i.e., $\mathcal{S}_k=\{1,2,\ldots, L_{k,k}-L_{\rm I}\}$. One may, however, construct a set of the transmit precoding vectors by selecting arbitrary $L_{\rm D}-L_{\rm I}$ column vectors from the IDFT matrix. Once the set of the precoding vectors is determined, all receivers exploit the decoding matrix that lies in the null space of the subspace spanned by the precoding vectors. As a result, with the predetermined decoding DFT matrix, all receivers can eliminate the aligned interference signals. Interesting future work would be to find the optimal $L_{\rm D}-L_{\rm I}$ column vectors (subcarriers) among $N$ column vectors in ${\bf F}$ that maximize the achievable sum-spectral efficiency by exploiting CSIT feedback.

 {\bf Remark 4 (Symmetric case):} Suppose the symmetric scenario in which $L_{k,k}=L_{\rm D}$ for $k\in\mathcal{K}$ and $L_{k,i}=L_{\rm I}$ for $k\neq i$ and $k,i\in\mathcal{K}$. In this case, the achievable sum-DoF simplifies to
\begin{align}
d^{\rm IC, Sym}_{\Sigma} = \begin{cases}
                        \frac{K}{2+\frac{L_{\rm I}-1}{L_{\rm D}-L_{\rm I}}} ~~~~~\text{for~~ $L_{\rm D} \geq \frac{3}{2}L_{\rm I}$} \\
                        \frac{K(L_{\rm D}-L_{\rm I})}{2L_{\rm I}-1} ~~~~~\text{for~~$L_{\rm I} < L_{\rm D} < \frac{3}{2}L_{\rm I}$}
                    \end{cases}. \label{eq:theorem2}
\end{align}
\eqref{eq:theorem2} shows that in the extreme case of $L_{\rm D} =2$ and $L_{\rm I}=1$, $d^{\rm IC, Sym}_{\Sigma} =\frac{K}{2}$ can be achieved. In another extreme case with  $L_{\rm D} =\frac{3}{2}L_{\rm I}-1$, one can achieve $d^{\rm IC, Sym}_{\Sigma}=\frac{K (\frac{1}{2}L_{\rm I}-1)}{2L_{\rm I}-1}$. These results imply that the sum-DoF converges to $d_{\Sigma}=\frac{K}{4}$ as $L_{\rm I}$ tends to infinity. Thus, we can expect the linear gain with respect to $K$ in the both cases. 

{\bf Remark 5 (Difference with blind interference alignment \cite{JafarBIA}):} 

The proposed method is similar to blind interference alignment because the both align the interference signals without using CSIT. Nevertheless, the proposed method differs from blind interference alignment in that it does not need a special channel coherence pattern to accomplish interference alignment. The only requirement for IF-ORDM is that the number of channel taps (i.e., the delay spread) be larger for the desired links than for the interfering links.

 {\bf Remark 6 (Comparison with TDMA-OFDM):} The proposed communication strategy resembles with the OFDM transmission method, which is widely used in the contemporary wireless systems (e.g. LTE and WiFi). For example, if each transmitter-and-receiver pair sends $M$ data symbols by using $M$-point IFFT operation with the addition of cyclic prefix length $\max\{L_{\rm I},L_{\rm D}\}-1$ in a round-robin fashion, the sum-DoF of the $K$-user interference with ISI is 
\begin{align}
d_{\Sigma}^{{\rm TDMA-OFDM}}=\frac{M}{M+\max\{L_{\rm I},L_{\rm D}\}-1}.
\end{align}
As $M$ goes infinity, it converges to unity. This trend implies that the OFDM method does not scale with the number of $K$.

The proposed transmission method can be interpreted as a variant of the conventional OFDM transmission, because IF-OFDM also uses IDFT to transform a set of data symbols. Nevertheless, the two methods differ. The key differences are: 1) the selection of cyclic prefix size and 2) the use of the subcarriers to remove IUI. IF-OFDM selects the cyclic prefix size to create the relativity of the channel structure, whereas OFDM chooses the size to create parallel subchannels of the desired link by removing ICI. Furthermore, the proposed method uses a set of column vectors of the IDFT matrix during transmit mode, and the receiver applies the non-overlapping column vectors of the DFT matrix to reject IUI. The conventional OFDM transmission, however, uses the same set of column vectors of IDFT and DFT matrices in transmitting and receiving mode.

\subsection{Examples}
It is instructive to consider an example to provide a better understanding our result.

 {\bf Example 3:} Suppose a symmetric case in which with $B=2$, $L_{k,k}=L_{\rm D}=3$, and $L_{k,i}=L_{{\rm I}}=1$; these parameters correspond to the scenario where two dominant multi-paths exist with the line-of-sight path in the desired channels, but no effective multi-paths are resolved in the interference channels. Because $L_{\rm I}=1$, no cyclic prefix is needed in this example. Furthermore, we set the size of each subblock (OFDM symbol) to ${\bar N}=2(L_{\rm D}-L_{\rm I})+L_{\rm I}-1=4$.

Because $L_{\rm D}-L_{\rm I}=2$, transmitter $k$ uses a 4-point IDTF matrix for precoding and sends two information symbols in during the $b$th subblock where $b=\{1,2\}$. In this example, suppose the active subcarrier set that carries  two data symbols in each subblock is $\mathcal{S}_k=\{1,3\}$.  Applying the first and third column vectors of the 4-point IDTF matrix as precoding vectors ${\bf f}_1= \frac{1}{\sqrt{4}} [1,  1, 1, 1]^{\top}$ and ${\bf f}_3= \frac{1}{\sqrt{4}} [1,  -1, 1, -1]^{\top}$, we create the channel input vector of each subblock as follows: 
\begin{align}
{\bf x}_k^{1}&= \left[%
\begin{array}{c}
{x}_{k}[1]   \\
{x}_{k}[2]  \\
{x}_{k}[3]   \\
{x}_{k}[4]  \\
\end{array}%
\right] =\frac{1}{\sqrt{4}} \left[%
\begin{array}{c}
1  \\
1  \\
1  \\
1  \\
\end{array}%
\right] s_{k,1}^1 +  \frac{1}{\sqrt{4}} \left[%
\begin{array}{c}
1  \\
-1  \\
1  \\
-1  \\
\end{array}%
\right] s_{k,3}^1\nonumber \\
{\bf x}_k^{2}&= \left[%
\begin{array}{c}
{x}_{k}[5]   \\
{x}_{k}[6]  \\
{x}_{k}[7]   \\
{x}_{k}[8]  \\
\end{array}%
\right] =\frac{1}{\sqrt{4}} \left[%
\begin{array}{c}
1  \\
1  \\
1  \\
1  \\
\end{array}%
\right] s_{k,1}^2 + \frac{1}{\sqrt{4}} \left[%
\begin{array}{c}
1  \\
-1  \\
1  \\
-1  \\
\end{array}%
\right] s_{k,3}^2 .
 \end{align}
Then, the receive signal at receiver $k\in\mathcal{K}$ until time slot 8 is given as in (\ref{eq:example2}).  
\begin{figure*}
\begin{align}\small
 \left[\!\!\!%
\begin{array}{c}
{y}_{k}[1]   \\
{y}_{k}[2]  \\
{y}_{k}[3]  \\
{y}_{k}[4]  \\
{y}_{k}[5]  \\
{y}_{k}[6]  \\
{y}_{k}[7]  \\
{y}_{k}[8]  \\
\end{array}%
\!\!\!\right] \!\!&\!=\! \!\small \left[\!\!\!%
\begin{array}{cccccccc}
{h}_{k,k}[1] \!\!&\!\!0   \!\!&\!\!0 \!\!&\!\!0 \!\!&\!\!0\!\!&\!\! 0\!\!&\!\!0\!\!&\!\! 0 \\
{h}_{k,k}[2]  \!\!&\!\! {h}_{k,k}[1] \!\!&\!\!  0 \!\!&\!\!0 \!\!&\!\!0 \!\!&\!\! 0 \!\!&\!\!0 \!\!&\!\! 0  \\
{h}_{k,k}[3] \!\!&\!\!{h}_{k,k}[2]\!\!&\!\!  {h}_{k,k}[1]  \!&\! 0  \!&\! 0 \!\!&\!\! 0 \!&\! 0 \!\!&\!\! 0\\
0 \!\!&\!\! {h}_{k,k}[3]\!\!&\!\!  {h}_{k,k}[2]  \!\!&\!\! {h}_{k,k}[1] \!\!&\!\! 0 \!\!&\!\!  0 \!\!&\!\! 0 \!\!&\!\!  0 \\
0 \!\!&\!\! 0\!\!&\!\! {h}_{k,k}[3] \!\!&\!\!{h}_{k,k}[2]   \!\!&\!\! {h}_{k,k}[1] \!&\!  0   \!\!&\!\! 0 \!&\!  0 \\
0  \!&\! 0 \!\!&\!\!0  \!\!&\!\!{h}_{k,k}[3] \!\!&\!\! {h}_{k,k}[2] \!\!&\!\!  {h}_{k,k}[1] \!\!&\!\! 0 \!\!&\!\! 0   \\
0 \!\!&\!\!0\!\!&\!\!  0\!\!&\!\! 0\!\!&\!\! {h}_{k,k}[3] \!\!&\!\!  {h}_{k,k}[2]  &\!\! {h}_{k,k}[1] \!\!&\!\!  0  \\
0 \!\!&\!\!0\!\!&\!\!  0\!\!&\!\! 0\!\!&\!\! 0\!\!&\!\!  {h}_{k,k}[3] \!\!&\!\!  {h}_{k,k}[2]  &\!\! {h}_{k,k}[1] \\
\end{array}%
\!\!\!\!\right]  \!\!\! \left[\!\!\!%
\begin{array}{c}
{x}_{k}[1]   \\
{x}_{k}[2]  \\
{x}_{k}[3]   \\
{x}_{k}[4]  \\
{x}_{k}[5]   \\
{x}_{k}[6]  \\
{x}_{k}[7]   \\
{x}_{k}[8]  \\
\end{array}%
\!\!\!\right]  \!+\! \sum_{\ell\neq k}^K h_{k,i}[1]{\bf I}_{8\times 8}\!\! \left[%
\begin{array}{c}
{x}_{i}[1]   \\
{x}_{i}[2]  \\
{x}_{i}[3]   \\
{x}_{i}[4]  \\
{x}_{i}[5]   \\
{x}_{i}[6]  \\
{x}_{i}[7]   \\
{x}_{i}[8]  \\
\end{array}%
\right]
\!+\! \small\left[%
\begin{array}{c}
{z}_{k}[1]   \\
{z}_{k}[2]  \\
{z}_{k}[3]  \\
{z}_{k}[4]   \\
{z}_{k}[5]  \\
{z}_{k}[6]  \\
{z}_{k}[7]  \\
{z}_{k}[8]  \\
\end{array}%
\right]. \label{eq:example2}
\end{align} 
\end{figure*}
In the first subblock, the effective input-output relationship is given by
\begin{align}
& \left[\!\!%
\begin{array}{c}
{y}_{k}[1]  \\
{y}_{k}[2]  \\
{y}_{k}[3]  \\
{y}_{k}[4]  \\
\end{array}%dd
\!\!\!\right] \!\!=  \left[\!\!%
\begin{array}{cccc}
{h}_{k,k}[1]  \!&\! 0&  0 \!&\! 0 \\
{h}_{k,k}[2]   \!&\! {h}_{k,k}[1]  & 0  \!&\! 0 \\
{h}_{k,k}[3]   \!&\! {h}_{k,k}[2]  & {h}_{k,k}[1]   \!&\! 0 \\
0  \!&\! {h}_{k,k}[3]  & {h}_{k,k}[2]   \!&\! {h}_{k,k}[1] \\
\end{array}%
\!\!\right]   \left[\!\!%
\begin{array}{c}
{x}_{k}[1]   \\
{x}_{k}[2]  \\
{x}_{k}[3]   \\
{x}_{k}[4]  \\
\end{array}%
\!\!\right] \nonumber \\\!&\!\!+\! \sum_{i\neq k}^K \left[\!\!%
\begin{array}{cccc}
{h}_{k,i}[1]  \!&\! 0 \!&\! 0 \!&\! 0 \\
 0\!&\! {h}_{k,i}[1]  \!&\! 0 \!&\! 0\\
0 \!&\! 0 \!&\! {h}_{k,i}[1] \!&\! 0 \\
 0\!&\! 0  \!&\! 0 \!&\! {h}_{k,i}[1]\\
\end{array}%
\!\!\right] \!\!\! \left[\!\!%
\begin{array}{c}
{x}_{i}[1]   \\
{x}_{i}[2]  \\
{x}_{i}[3]   \\
{x}_{i}[4]  \\
\end{array}%
\!\!\right] \!\!+\!\! \left[\!\!%
\begin{array}{c}
{z}_{k}[1]   \\
{z}_{k}[2]  \\
{z}_{k}[3]   \\
{z}_{k}[4]  \\
\end{array}%
\!\!\!\right] \label{eq:simple1}\!\!.
\end{align} 
Since ${\bf x}^1_k ={\bf f}_1s_{k,1}^1+{\bf f}_3s_{k,3}^1$ for $k\in\mathcal{K}$, \eqref{eq:simple1} is can be written as 
\begin{align}
 \left[%
\begin{array}{c}
{y}_{k}[1]  \\
{y}_{k}[2]  \\
{y}_{k}[3]  \\
{y}_{k}[4]  \\
\end{array}%dd
\right] \!&\!= {\bf H}_{k,k}( {\bf f}_1 s_{k,1}^1 +{\bf f}_3 s_{k,3}^1) \nonumber \\\!&\!+ \sum_{i \neq k}^K {h}_{k,i}[1]( {\bf f}_1s_{i,1}^1+{\bf f}_3s_{i,3}^1)+ \left[%
\begin{array}{c}
{z}_{k}[1]   \\
{z}_{k}[2]  \\
{z}_{k}[3]   \\
{z}_{k}[4]  \\
\end{array}%
\right]\label{eq:simple2}.
\end{align}
By this procedure, all the interference signals are aligned in the direction of ${\bf f}_1$ and ${\bf f}_3$, whereas the desired signal is not aligned to the interference subspace. By multiplying ${\bf f}_2=\frac{1}{\sqrt{4}}[1,  -j , -1, j]^{\top}$ and ${\bf f}_4=\frac{1}{\sqrt{4}}[1,  j , -1, -j]^{\top}$ to ${\bf y}_k^1=[y_k[1],~ y_k[2],~ y_k[3],~ y_k[4]]^{\top}$, the aligned IUI signals are canceled, and we finally obtain the output:
\begin{align}
 {\bf \tilde y}^1_k= \underbrace{\left[\!\!%
\begin{array}{cc}
{\bf f}_2^{\top}{\bf  H}_{k,k}{\bf f}_1 & {\bf f}_2^{\top}{\bf  H}_{k,k}{\bf f}_3 \\
{\bf f}_4^{\top}{\bf  H}_{k,k}{\bf f}_1& {\bf f}_4^{\top}{\bf  H}_{k,k}{\bf f}_3\\
\end{array}%dd
\!\!\!\right] }_{{\bf \tilde H}_k} \left[\!\!%
\begin{array}{c}
s^1_{k,1}  \\
s^1_{k,3}  \\
\end{array}%dd
\!\!\!\right] +  \left[\!\!%
\begin{array}{c}
{\bf f}_2^{\top}{\bf  z}_k\\
{\bf f}_4^{\top}{\bf  z}_k\\
\end{array}%dd
\!\!\!\right]. 
\end{align}
Since ${\rm rank}\left({\bf \tilde H}_k\right)=2$ with high probability, $s_{k,1}^1$ and $s_{k,3}^1$ can be reliably decoded by using the ZF decoder to eliminating ICI.

 We now consider the received signals in the second subblock, which are written as the superposition of the desired signal, the inter-subblock interference, and the IUI. The effective input-output relationship of the $b$th block is given by
\begin{align}
& \left[\!\!%
\begin{array}{c}
{y}_{k}[5]  \\
{y}_{k}[6]  \\
{y}_{k}[7]  \\
{y}_{k}[8]  \\
\end{array}%dd
\!\!\!\right]  =  \underbrace{{\bf H}_{k,k} ({\bf f}_1s_{k,1}^2 +{\bf f}_3s_{k,3}^2)}_{{\rm desired~ signal}}+ \underbrace{\sum_{i \neq k}^K {h}_{k,i}[1]( {\bf f}_1s_{i,1}^2+{\bf f}_3s_{i,3}^2)}_{{\rm inter-cell~interference}} \nonumber \nonumber \\& +  \underbrace{\left[\!\!%
\begin{array}{cccc}
0  \!&\! 0&   {h}_{k,k}[3] \!&\!  {h}_{k,k}[2] \\
0   \!&\! 0& 0  \!&\!  {h}_{k,k}[3] \\
0  \!&\! 0  & 0   \!&\! 0 \\
0  \!&\!0  & 0  \!&\!0 \\
\end{array}%
\!\!\right] \left[\!\!%
\begin{array}{c}
{x}_{k}[1]   \\
{x}_{k}[2]  \\
{x}_{k}[3]   \\
{x}_{k}[4]  \\
\end{array}%
\!\!\right]}_{{\rm inter-subblock ~interference}} +\left[\!\!%
\begin{array}{c}
{z}_{k}[1]   \\
{z}_{k}[2]  \\
{z}_{k}[3]   \\
{z}_{k}[4]  \\
\end{array}%
\!\!\!\right],
\end{align}
This equation shows that inter-subblock interference terms ${h}_{k,k}[3]{x}_{k}[3]+{h}_{k,k}[2]{x}_{k}[4]$ and ${h}_{k,k}[3]{x}_{k}[4]$ can be cancelled, as receiver $k\in\mathcal{K}$ has decoded these information symbols in the previous subblock. Accordingly, the same input-output relationship after eliminating the inter-subblock interference is given by
\begin{align}
& \left[%
\begin{array}{c}
{y}_{k}[5]  \\
{y}_{k}[6]  \\
{y}_{k}[7]  \\
{y}_{k}[8]  \\
\end{array}%dd
\right] -\left[\!\!%
\begin{array}{cccc}
0  \!&\! 0&   {h}_{k,k}[3] \!&\!  {h}_{k,k}[2] \\
0   \!&\! 0& 0  \!&\!  {h}_{k,k}[3] \\
0  \!&\! 0  & 0   \!&\! 0 \\
0  \!&\!0  & 0  \!&\!0 \\
\end{array}%
\!\!\right] \left[\!\!%
\begin{array}{c}
{x}_{k}[1]   \\
{x}_{k}[2]  \\
{x}_{k}[3]   \\
{x}_{k}[4]  \\
\end{array}%
\!\!\right] \nonumber \\
 \!&\!= {\bf H}_{k,k}( {\bf f}_1 s_{k,1}^2 +{\bf f}_3 s_{k,3}^2) \nonumber \\\!&\!+ \sum_{i \neq k}^K {h}_{k,i}[1]( {\bf f}_1s_{i,2}^2+{\bf f}_3s_{i,2}^2)+ \left[%
\begin{array}{c}
{z}_{k}[1]   \\
{z}_{k}[2]  \\
{z}_{k}[3]   \\
{z}_{k}[4]  \\
\end{array}%
\right].\end{align}
Applying the same decoding strategy as in the first subblock,  $s_{k,1}^2$ and $s_{k,3}^2$ can also be decoded. As a result, receiver $k$ can decode the four independent data symbols $\left\{s_{k,1}^1,s_{k,3}^1,s_{k,1}^2,s_{k,3}^2\right\}$ with the ten channel uses.

\begin{figure*}
\centering \vspace{-0.1cm}
\includegraphics[width=6.6in]{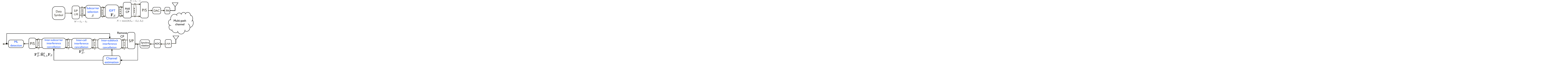} \vspace{-0.1cm}\caption{One possible transmitter and receiver architecture for IF-OFDM.} \label{fig:2} \vspace{-0.3cm}
\end{figure*}

When $B$ subblocks are used, the achievable rate of the $k$th link is 
\begin{align}
R_k = \frac{B}{4B+2}\log_2\left( \det\left({\bf I}_{2\times 2} + \frac{P }{2\sigma^2}{\bf \tilde H}_k{\bf \tilde H}_k^H\right)\right).
\end{align}
Consequently, as $B$ goes to infinity, the achievable sum-spectral efficiency of this channel is asymptotically given by
\begin{align}
\lim_{B\rightarrow \infty}\sum_{k\in\mathcal{K}}R_k = \sum_{k\in\mathcal{K}}\frac{1}{4}\log_2\left( \det\left({\bf I}_{2\times 2} + \frac{P }{2\sigma^2}{\bf \tilde H}_k{\bf \tilde H}_k^H\right)\right).
\end{align}

This example demonstrates that the proposed communication strategy achieves the sum-DoF of the channel without CSIT, $d^{\star}_{\Sigma}=\frac{K}{2+\frac{L_{\rm I}-1}{L_{\rm D}-L_{\rm I}}}=\frac{K}{2}$. Fig. \ref{fig:2} illustrates one possible transmitter and receiver architecture to implement the proposed method explained in this example.

 %Another interesting observation here is that $R_k$ for $k\in\mathcal{K}$ is a function of ${\rm SNR}$ and the multi-path channel gain $|h_{k,k}[2]|^2$. In other words, the LOS channel gain $|h_{k,k}[2]|^2$ is irrelevant to the transmission rate. Therefore, the achievable rate can improve in the communication scenario where the multi-path channel gain is large. 

 {\bf Remark 7 (Inter-subblock Interference Cancellation):} One may implement the proposed method without using inter-subblock interference cancellation operation by including $L_{\rm D}-L_{\rm I}$ guard-time (zero) slots with the cyclic prefix between consecutive subblocks. This can reduce the computational complexity of interference cancellation algorithms at the cost of the increased underutilization of the resources.

\section{Characterization of Sum-Spectral Efficiency} 

In this section, we characterize the achievable sum-spectral efficiency of the SISO frequency-selective interference channel with 1) complete lack of CSIT and 2) the desired link's CSIT.

\subsection{ With Completely No CSIT }

We first provide a characterization of the sum-spectral efficiency by using the proposed method explained in the previous section, which requires no CSIT. From \eqref{eq:concentrating_rx_signals}, the output signal vector of the $b$th subblock at receiver $k$ after canceling the inter-subblock interference is given by
\begin{align}
{\bf \tilde y}_k^b -{\bf F}_{\mathcal{S}_k^c}^H{\bf \bar H}^{b,b-1}_{k,k}{\bf F}_{\mathcal{S}_k}{\bf s}_k^{b-1} = {\bf F}_{\mathcal{S}_k^c}^H{\bf \bar H}^b_{k,k}{\bf F}_{\mathcal{S}_k}{\bf s}_k^b+{\bf \tilde z}^b_k.
\end{align}
Applying the QR decomposition \cite{Golub:96}, the effective channel matrix $ {\bf F}_{\mathcal{S}_k^c}^H{\bf \bar H}_{k,k}^b{\bf F}_{\mathcal{S}_k}$ can be represented by a product of a unitary matrix ${\bf Q}_{k,k}^b\in\mathbb{C}^{(L_{k,k}-L_{\rm I}) \times (L_{k,k}-L_{\rm I}) }$ and an upper-triangular matrix ${\bf R}^b_{k,k}\in\mathbb{C}^{(L_{k,k}-L_{\rm I}) \times (L_{k,k}-L_{\rm I})}$,
namely, 
\begin{align}
 {\bf F}_{\mathcal{S}_k^c}^H{\bf \bar H}_{k,k}^b{\bf F}_{\mathcal{S}_k}= {\bf Q}^b_{k,k}{\bf R}^b_{k,k}.
\end{align}
Under the premise that each transmitter knows the effective SNR and uses adaptive modulation/coding to select the right rate, the achievable rate of the $k$th message $m_k$ sent over $M=B(N+L_{\rm I}-1)+L_{\rm D}-1$ time slots by ZF-SIC is computed as
\begin{align}
R_k= \frac{\sum_{n=1}^{L_{k,k}-L_{\rm I}}\sum_{b=1}^{B} \log_2\left(1+|r_{k,k}^{n,b}|^2{\rm SNR}\right)}{B(N+L_{\rm I}-1)+L_{\rm D}-1},
\end{align}
where $r_{k,k}^{n,b}$ is the $n$th diagonal element of ${\bf R}_{k,k}^b$ and ${\rm SNR}=\frac{\frac{N}{N+L_{\rm I}+1}\frac{NP}{L_{k,k}-L_{\rm I}}}{\sigma^2}$ because a fraction $\frac{N}{N+L_{\rm I}+1}$ of power is only used for the data transmission with $\mathbb{E}\left[|s_{k}^{b,n}|^2\right]=\frac{NP}{L_{k,k}-L_{\rm I}}$ to satisfy $\mathbb{E}[\|{\bf F}_{\mathcal{S}_k}{\bf s}_k^{b} \|^2]=NP$. Notice that $r_{k,k}^{n,b}=r_{k,k}^{n,q}$ for all $b,q\in\{1,2,\ldots,B\}$ because the channel coefficients in ${\bf \bar H}_{k,k}^b$ are invariant over different subblocks. By symmetry, the achievable sum-spectral efficiency is \begin{align}
\lim_{B\rightarrow \infty}\sum_{k=1}^KR_k = \! \frac{\sum_{k=1}^{K}\!\sum_{n=1}^{L_{k,k}-L_{\rm I}} \log_2\left(\!1\!+\!|r_{k,k}^{n,1}|^2{\rm SNR}\right)}{N+L_{\rm I}-1}, \label{eq:rates_with_no_CSIT}
\end{align}
as $B$ tends to infinity. 

\begin{figure}
\centering \vspace{-0.3cm}
\includegraphics[width=3.5in]{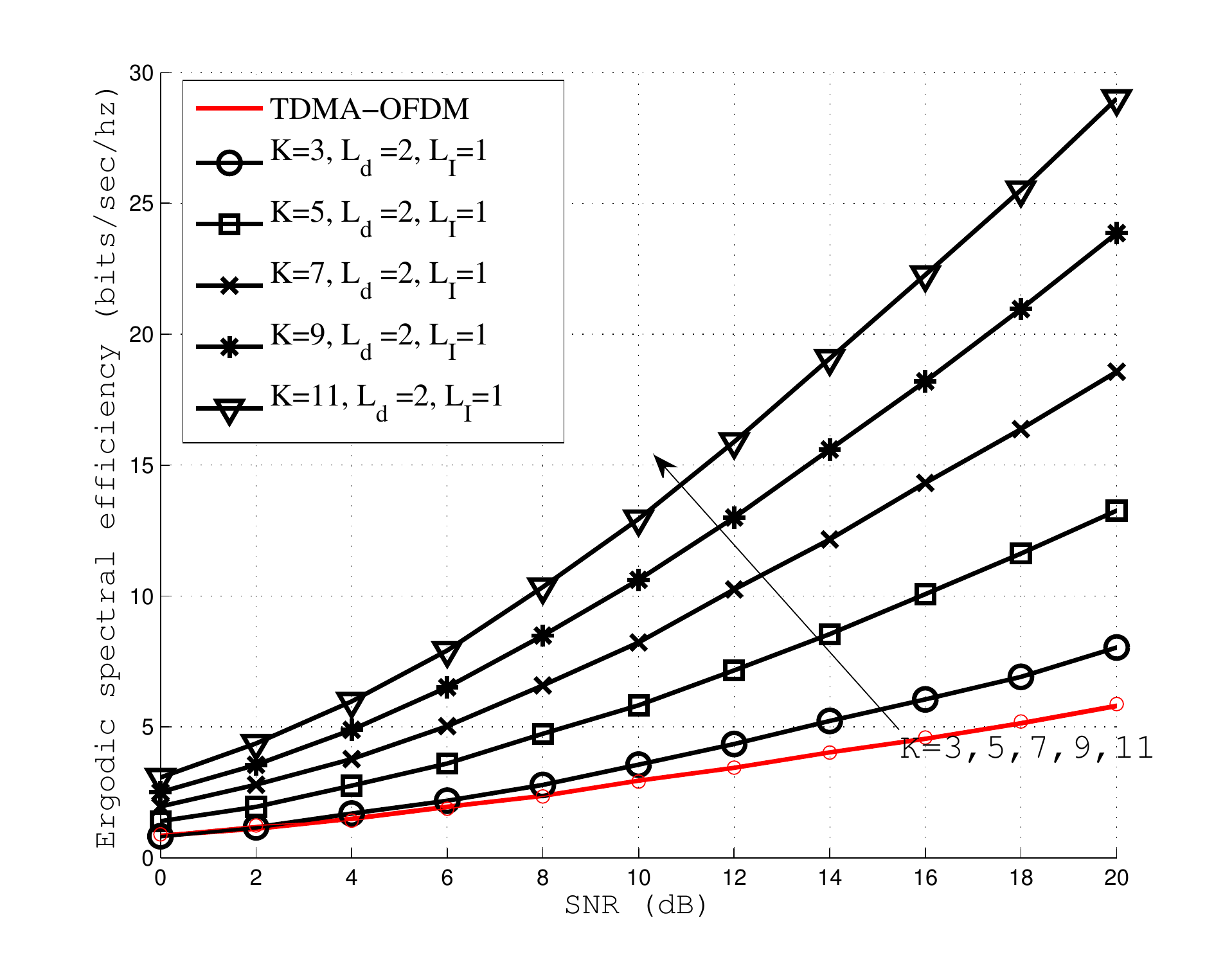} \vspace{-0.7cm}\caption{The ergodic sum-spectral efficiency for different $K$ when $L_{k,k}=L_{\rm D}=2$ and $L_{k,i}=L_{\rm I}=1$, in which $h_{k,i}[\ell]$ is generated from $\mathcal{CN}(0,1)$ for all $i,j,\ell$. } \label{fig:3} 
\end{figure}

Fig. \ref{fig:3} shows that the ergodic sum-spectral efficiency increases linearly with $K$ when $L_{k,k}=L_{\rm D}=2$ and $L_{k,i}=L_{\rm I}=1$. It is also notable that IF-OFDM outperforms TDMA-OFDM in all SNR regimes.  
 
\subsection{With CSIT of the Desired Link}

We now consider the case when CSIT of the desired link is available, i.e., ${\bf \tilde H}^b_{k,k}={\bf F}_{\mathcal{S}_k^c}^H{\bf \bar H}^b_{k,k}{\bf F}_{\mathcal{S}_k}$ is known to transmitter $k$. Under this premise, we provide a characterization of the achievable sum-spectral efficiency by using the proposed two-stage precoding and decoding method with the optimal power allocation via water-filling. 

It is well-known that the effective channel matrix ${\bf \tilde H}^b_{k,k}={\bf F}_{\mathcal{S}_k^c}^H{\bf \bar H}^b_{k,k}{\bf F}_{\mathcal{S}_k}$ can be diagonalized by two unitary matrices ${\bf U}_{k,k}\in\mathbb{C}^{(L_{k,k}-L_{\rm I})\times (L_{k,k}-L_{\rm I})}$ and ${\bf V}_{k,k}\in\mathbb{C}^{(L_{k,k}-L_{\rm I})\times (L_{k,k}-L_{\rm I})}$ by using SVD \cite{Golub:96},
namely, 
\begin{align}
{\bf \tilde H}_{k,k}^b={\bf F}_{\mathcal{S}_k^c}^H{\bf \bar H}_{k,k}^b{\bf F}_{\mathcal{S}_k}= {\bf U}_{k,k}^b{\bf \Sigma}_{k,k}^b{{\bf V}_{k,k}^b}^H,
\end{align}
where ${\bf \Sigma}_{k,k}^b={\rm diag}\left(\left[{\tilde h}_{k,k}^{b,1},{\tilde h}_{k,k}^{b,2}, \ldots, {\tilde h}_{k,k}^{b,L_{k,k}\!-\!L_{\rm I}}\right]\right)$ is a diagonal matrix that consists of the singular values of effective channel matrix ${\bf \tilde H}_{k,k}^b$. From this decomposition, we use the proposed two-stage precoding method to construct transmit signal vector ${\bf x}_k$ as
\begin{align}
{\bf \bar x}_k= {\bf F}_{\mathcal{S}_k}{\bf V}^b_{k,k}{\bf P}^b_{k,k}{\bf s}_k^b,
\end{align}
where ${\bf V}^b_{k,k}$ is the inner precoding matrix that creates multiple parallel subchannels by eliminating IUI; ${\bf F}_{\mathcal{S}_k}$ is the outer precoding matrix that aligns inter-user-interference. ${\bf P}_{k,k}^b ={\rm diag}\left(\left[\sqrt{p_{k,k}^{b,1}}, \ldots, \sqrt{ p_{k,k}^{b,L_{k,k}-L_{\rm I}}}\right]\right)$ is the diagonal matrix in which the $n$th diagonal element $p_{k,k}^{b,n}$ is power allocated to data symbol $s_{k,n}^{b}$. Since this two-stage precoding method does not change the subspace occupied by the IUI signals, the IUI signals are removed by multiplying ${\bf F}_{\mathcal{S}_k^c}^H$ to the received vector ${\bf \bar y}_k$ after discarding the cyclic prefix. Multiplying ${{\bf U}_{k,k}^b}^{\!\!\!\!\!\!H}$ to ${\bf \tilde y}_k$, we have $L_{k,k}-L_{\rm I}$ parallel sub-channel output vector:
\begin{align}
{\bf \hat y}_k^b \!&\!= {{\bf U}_{k,k}^b}^{\!\!\!\!\!\!H}  {\bf \tilde H}_{k,k}^b{\bf V}_{k,k}^b{\bf s}_{k}^b+ {{\bf U}_{k,k}^b}^{\!\!\!\!\!\!H}{\bf \tilde z}_k^b
\nonumber \\\!&\!= {\bf \Sigma}_{k,k}^b{\bf s}_{k}^b + {\bf \hat z}_k^b.
\end{align}  
Then, assuming that a separate
capacity-achieving AWGN code is used to communicate over each of parallel subchannels, 
the maximum sum-spectral efficiency achievable using this scheme for large enough $B$ is
\begin{align}
\lim_{B\rightarrow \infty}\sum_{k=1}^KR_k\!=\! \frac{ \! \sum_{k=1}^K\!\sum_{n=1}^{L_{k,k}\!-\!L_{\rm I}}\log_2\!\left(\!1\!+\!\frac{ |{\tilde h}_{k,k}^{b,n}|^2{\bar p}^{b,n}_{k,k}}{\sigma^2}\!\right)}{ N+L_{\rm I}-1},
 %\nonumber \\
%\!&\!{\rm subject~~ to}~ \sum_{m=1}^{N/2} p^{\star}_{i,m} \leq \frac{NP}{2}~~ {\rm and}~~~ p_{i,m}\geq 0.
\end{align}
where ${\bar p}^{b,n}_{k,k}$ is the optimal power allocated to the $n$th subchannel based on \textit{water-filling strategy} in 
\cite{Hirt_Massey:88}, namely,
\begin{align}
{\bar p}^{b,n}_{k,k}= \left(\frac{1}{\delta}-\frac{\sigma^2}{ |{\tilde h}_{k,k}^{b,n}|^2}\right)^{+},
\end{align}
where $\delta$ is the parameter associated with a Lagrange multiplier that is selected to satisfy the power constraint.

\begin{figure}
\centering \vspace{-0.1cm}
\includegraphics[width=3.3in]{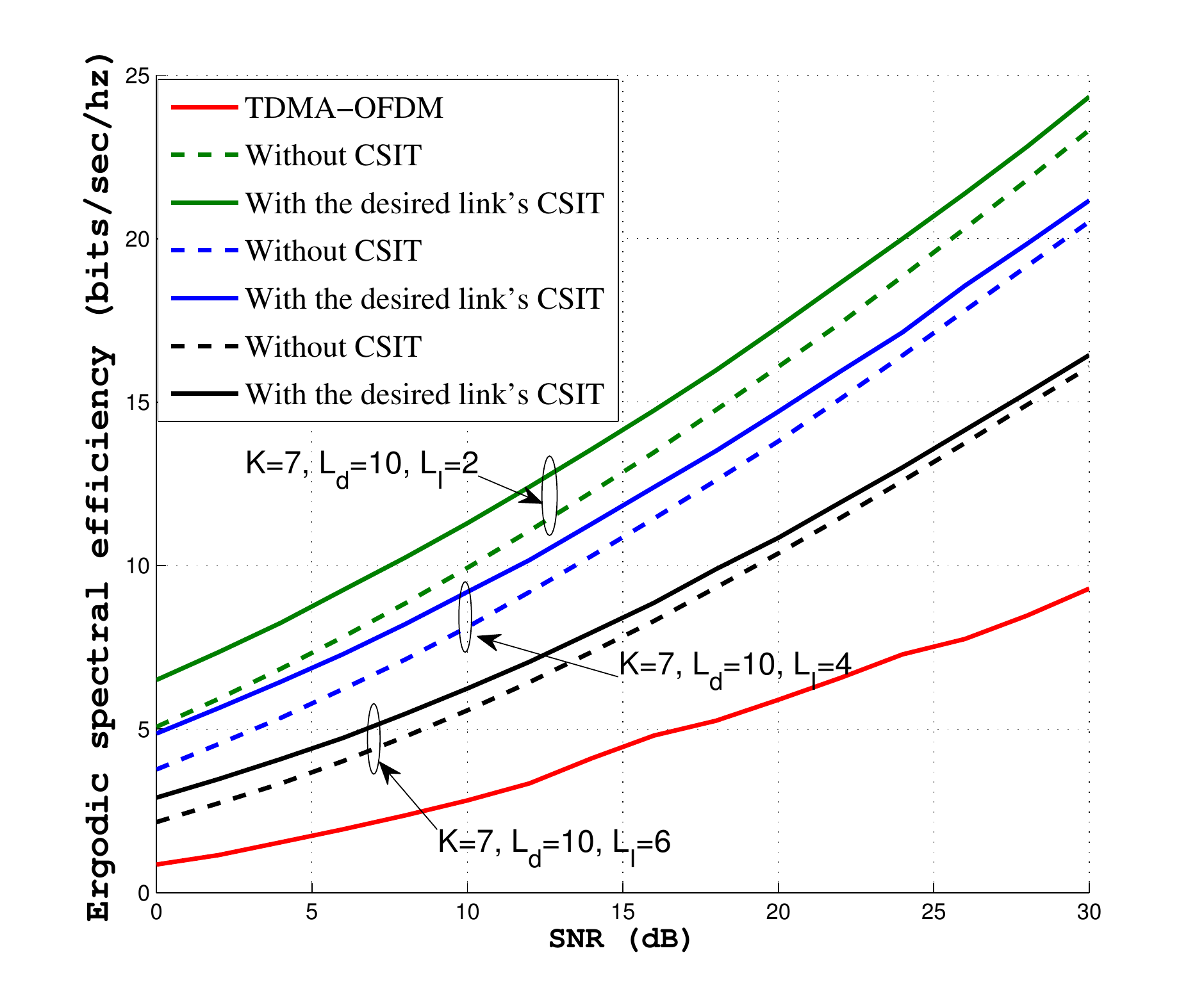} \vspace{-0.5cm}\caption{ The ergodic sum-spectral efficiency with the desired link's CSIT. In the simulation, $h_{k,i}[\ell]$ is generated from $\mathcal{CN}(0,1)$ for all $i,j,\ell$. } \label{fig:4} \vspace{-0.1cm}
\end{figure}

This two-stage transmission method provides a better achievable sum-spectral efficiency than the case with completely no CSIT, whereas the same sum-DoF achieves with it. Fig. \ref{fig:4} illustrates the ergodic sum-spectral efficiency when $K=7$, $L_{k,k}=L_{\rm D}=10$, and $L_{k,i}=L_{\rm I}\in\{2,4,6\}$. As can be seen, CSIT feedback for the direct link provides the gain in terms of the ergodic sum-spectral efficiency in the low SNR regime when $L_{\rm I}=6$. For $L_{\rm I}\in\{2,4\}$, the proposed two-stage transmission can substantially improve the sum-spectral efficiency, even in the high SNR regime.

{\bf Remark 8 (Separability):} Our two-stage precoding communication strategy is practically relevant because it provides a significant spectral efficiency gain in a multi-carrier interference channel with separate encoding over each carrier and a power allocation across carriers.

%%%%%%%%%%%%%%%%%%%%%%%%%%%%%%%%%%%%%%%%%%%%%%%%%%%%%%%%%%%%%%%%%%
 \section{Discussion}

The fundamental principle of IF-OFDM is to exploit the circulant matrix property so that IUI is aligned in the subspace spanned by a certain set of column vectors of a Fourier matrix. This principle can be applicable in numerous communication scenarios. In this section, we first explain some scenarios in which the proposed method shows a better sum-DoF gain in interference networks when CSIT is not available. In the sequel, we discuss some practical ways how to apply the proposed method when $L_{\rm D} < L_{\rm I}$.

\subsection{Extension to Other Interference Networks}
In this subsection, we consider two channel models: 1) multiple-input-single-output (MISO) broadcast channel with ISI and 2) multiple-input-multiple-output (MIMO) frequency-flat $K$-user interference channel with a circulant structure.

\subsubsection{MISO broadcast channel with ISI}  One interesting observation is that the proposed method is directly applicable to a MISO broadcast channel with ISI when CSIT is not available. Suppose that $K$ transmitters cooperate to  form a virtual transmitter with $K$ distributed antennas such as cloud radio access networks (C-RAN) \cite{C-RAN}. In this case, the proposed method enables simultaneous support for the $K$ users, even with completely no CSIT, provided that the channel condition of the scheduled user set satisfies the condition of $L_{k,k}> L_{\rm I}$. In what follows, we present this claim precisely.

{\bf Corollary 1:} Suppose a MISO broadcast channel with ISI in which a transmitter equipped with $K$ distributed antennas communicates with $K$ users, each with a single antenna. Let $L_{k,i}$ be the number of channel-taps from the $i$th transmit antenna to the $k$th user. With completely no CSIT, the sum-DoF of this channel is lower bounded by
\begin{align}
d^{\rm MISO-BC}_{\Sigma} \geq d^{\rm IC}_{\Sigma}.
\end{align}

\proof Since the cooperation of $K$ antennas at the transmitter does not degrade the sum-DoF, Corollary 1 holds from Theorem \ref{Theorem1}.
\endproof
 %
%This corollary demonstrates that ISI can help in obtaining the sum-DoF gain in a vector broadcast channel with completely no CSIT. 

\subsubsection{MIMO Interference Channel} Consider a frequency-flat (narrowband) MIMO $K$-user interference channel in which all transmitters and receivers have $N$ antennas. Let ${\bf G}_{k,i} \in\mathbb{C}^{N\times N}$ be the channel matrix from transmitter $i$ to receiver $k$. The following corollary shows the sum-DoF of the MIMO interference channel with a special channel structure.

{\bf Corollary 2:}Suppose ${\bf G}_{k,i}$ be a circulant matrix for $i\neq k$ and ${\bf G}_{k,k}$ be a non-circulant matrix composed of elements drawn from a continuous distribution. Then, when CSIT is not available, the proposed method achieves the sum-DoF of this channel as  
\begin{align}
d^{\rm MIMO-IC}_{\Sigma} = \frac{KM}{2}.
\end{align}

\proof The proof is direct from Theorem \ref{Theorem1}.
\endproof

\subsection{Treating Partial ISI as Noise When $L_{\rm I} > L_{\rm D}$ }

%One major concern of the proposed method is the requirement  that the number of channel-taps (or delay spreads) of desired link is lager than that of interfering links. In practice, finding the communication environment satisfying always this condition may be challenging. Even in the case of $L_{\rm I} > L_{\rm d}$, however, one can still apply the proposed method by treating a fraction of IUI as additional noise.

When $L_{\rm I} > L_{\rm D}$, one can also apply the proposed idea by treating a fraction of ISI as additional noise. This approach is of interest to the case when a distance-dependent large-scale model is incorporated in multi-cell wireless systems. 

{\bf Example 4:} Let $d_{k,i}$ be the distance between transmitter $i$ to receiver $k$ in the network. Then, the received signal of receiver $k$ can be rewritten by incorporating the distance-dependent large-scale model as 
\begin{align}
y_{k}[n]&=d_{k,k}^{-\frac{\alpha}{2}} \left(\sum_{\ell=1}^{L_{d}} h_{k,k}[\ell]x_{k}[n-\ell+1]\right) \nonumber \\
& + \sum_{i\neq k}^K d_{k,i}^{-\frac{\alpha}{2}} \left(\sum_{\ell=1}^{L_{\rm I}}h_{k,i}[\ell]x_{i}[n-\ell+1]\right) \!+\! {z}_k[n], 
\end{align}
where $\alpha$ is the path-loss exponent. When each receiver is in the cell center area, i.e., $d_{k,k} < d_{k,i}$, receiver $k$ can ignore the last ISI terms in $h_{k,i}[\ell]x_i[n-\ell]$ for $\ell \in\{L_{\rm D},L_{\rm D}+1,\ldots, L_{\rm I}\}$ by treating them as the additional noise, i.e.,
\begin{align}
d_{k,k}^{\frac{\alpha}{2}}y_{k}[n]&= \left(\sum_{\ell=1}^{L_{d}} h_{k,k}[\ell]x_{k}[n-\ell+1]\right) \nonumber \\
& + \sum_{i\neq k}^K\! \left(\frac{d_{k,k}}{d_{k,i}}\right)^{\!\!\!-\frac{\alpha}{2}}\!\! \left(\sum_{\ell=1}^{L_{d}-1}h_{k,i}[\ell]x_{i}[n-\ell+1]\right) \!+\! {z}^{{\rm eff}}_k[n], \nonumber
\end{align}
where
\begin{align}
{z}^{{\rm eff}}_k[n]= \sum_{i\neq k}^K \left(\frac{d_{k,k}}{d_{k,i}}\right)^{\!\!\!-\frac{\alpha}{2}}\sum_{\ell=L_d}^{L_{I}}h_{k,i}[\ell]x_{i}[n-\ell+1]+ d_{k,k}^{\frac{\alpha}{2}}z_k[n]. \nonumber
\end{align}
Assuming that $h_{k,i}[\ell]$ is statistically independent for all different $i,j,\ell$, and its power decays exponentially as the number of channel-taps increases, i.e., $\mathbb{E}[|h_{k,i}[\ell]|^2]=e^{-\beta (\ell-1)}$, the variance of the effective noise is
\begin{align}
\mathbb{E}\left[|{z}^{{\rm eff}}_k[n]|^2\right] = \sum_{i\neq k}^K\left(\frac{d_{k,k}}{d_{k,i}}\right)^{\!\!\!-\alpha} \sum_{\ell=L_d}^{L_{I}}   e^{-\beta (\ell-1)} P + d_{k,k}^{\alpha}\sigma^2,
\end{align}
This effective noise is negligible when $d_{k,k} \ll d_{k,i}$, so
the use of the proposed idea in multi-cell systems can still guarantee a reasonable performance. This example has had 
the purpose of showing the feasibility of the proposed method when $L_{\rm D} < L_{\rm I}$ by incorporating a large-scale fading effect. More rigorous system-level performance evaluation should be conducted to gauge the practical gain; we leave this for future work.
 
%%%%%%%%%%%%%%%%%%%%%%%%%%%%%%%%%%%%%%%%%%%%%%%%%%%%%%%%%%%%%%%%%%
 
  \section{Conclusion}
 We have demonstrated that the sum-DoF of the SISO $K$-user interference channel with ISI can increase linearly with $K$, even when CSIT is totally absent. This result is surprising because, in the SISO interference channel even without ISI, this linear sum-DoF gain with respect to $K$ is achievable when global and perfect CSIT are available across all transmitters or special channel patterns occur; this requirement has been a major obstacle to implementing interference alignment in practice. Our main result is proven by introducing a novel communication strategy, IF-OFDM. The main idea of the scheme is to create relativity of alignment based on different matrix structures, i.e., circulant and non-circulant structures, which are not subject to the realization of the channel coefficients.

A promising direction for future work is to consider a multi-antenna setting, (e.g., a multiple-input-multiple-output (MIMO) $K$-user interference channel with ISI), to determine whether the sum-DoF can increase with the both the numbers of antennas and of users when CSIT is unavailable. Other directions for future work include a study of system-level performance for our proposed scheme in cooperative cellular networks. In particular, by using a stochastic geometry framework \cite{Andrews_Tractable}, one can obtain a closed-form expression of the SINR distribution and gauge the gains in terms of spatially-averaged ergodic spectral efficiency to revisit the fundamental limit of cooperation \cite{Lozano_andrew_heath,Lee_David_Lozano_heath,Park}. Furthermore, one could seek ways to exploit the relativity of alignment based on circulant matrix property to design a new communication strategy in numerous multi-user communication scenarios when CSIT is not available. 
Lastly, a derivation of the sum-DoF outer bounds for the $K$-user interference channel with ISI would be useful from an information-theoretic perspective.

\end{document}